\documentclass[11pt]{JHEP3}

\usepackage{amsmath}
\usepackage{epsfig}
\usepackage{amssymb,amsfonts}

\usepackage{multirow,cite}
\usepackage{datetime}
\usepackage{graphics}

\newcommand{\bi}{\begin{itemize}}
\newcommand{\ei}{\end{itemize}}
\newcommand{\non}{\nonumber}
\def\half{\frac{1}{2}}
\def\p{\partial}
\def\a{\alpha}
\def\b{\beta}
\def\d{\delta}

\def\l{\lambda}

\def\om{\omega}
\def\e{\epsilon}
\def\s{\sigma}
\def\r{\rightarrow}

\def\k{\kappa}

\def\cL{\mathcal{L}}
\def\X{\mathrm{X}}
\def\cJ{\mathcal{J}}
\def\O{\mathcal{O}}

\newcommand{\bea}{\begin{eqnarray}}
\newcommand{\eea}{\end{eqnarray}}
\newcommand{\be}{\begin{equation}}
\newcommand{\ee}{\end{equation}}
\def\nn{\nonumber}
\newcommand{\eps}{\epsilon}

\title{Two Virasoro symmetries in stringy warped AdS$_3$
}
\preprint{\today, \currenttime}
\author{Geoffrey Comp\`{e}re$^\diamond$, Monica Guica$^\dag$ and Maria J. Rodriguez$^\sharp$ $^   \flat$\\

\vspace{1mm}
\hspace{-4.5mm}{\small $ {}^\diamond$    Physique Th\'eorique et Math\'ematique,\\
\hspace{-0.35 cm}
Universit\'e Libre de Bruxelles and International Solvay Institutes\\
\hspace{-0.35 cm}
Campus Plaine C.P.\ 231, B-1050 Bruxelles, Belgium\\
}

\hspace{-4.5mm}{\small $ {}^\dag$   David Rittenhouse Laboratory, University of Pennsylvania, \\  \hspace{-0.35 cm}  Philadelphia, PA 19104-6396, USA\\
}

\hspace{-4.5mm}{\small $ {}^ \sharp$   Center for the Fundamental Laws of Nature, Harvard University,\\ \hspace{-0.35 cm} Cambridge, MA 02138, USA.\\
}

\hspace{-4.5mm}{\small $ {}^    \flat$  Institut de Physique Th\'eorique, CEA Saclay, \\ \hspace{-0.35 cm} CNRS URA 2306 , F-91191 Gif-sur-Yvette, France.}}

\abstract{

\vspace{0.7cm}

We study three-dimensional consistent truncations of type IIB supergravity which admit warped AdS$_3$ solutions. These theories contain subsectors that have no bulk dynamics.  We show that the symplectic form for these theories, when restricted to the non-dynamical subsectors, equals the symplectic form for pure Einstein gravity in AdS$_3$. Consequently, for each consistent choice of boundary conditions in AdS$_3$, we can define a consistent phase space in warped AdS$_3$ with identical conserved charges. This way,  we easily obtain a Virasoro $\times$ Virasoro  asymptotic symmetry algebra in warped AdS$_3$; two different types of Virasoro $\times$ Ka\v{c}-Moody symmetries are also consistent alternatives.

\hspace{0.5 cm} Next, we study the phase space of these theories when propagating modes are included. We show that, as long as one can define a conserved symplectic form without introducing instabilities, the Virasoro  $\times$ Virasoro asymptotic symmetries can be extended to the entire (linearized) phase space. This implies that, at least at semi-classical level,
consistent theories of gravity in warped AdS$_3$  are described by a two-dimensional conformal field theory,  as long as  stability is not an issue.

}


\begin{document}

\section{Introduction}

Black holes carry an entropy given by a remarkably simple, yet completely universal, formula: the area of their event horizon in Planck units, divided by four \cite{Bekenstein:1973ur,Bardeen:1973gs}. Despite many years of investigation, the microscopic origin of this formula and the reason behind its universality have only been understood for a very special class of black holes, namely those whose near-horizon region contains an AdS$_3$ factor \cite{Strominger:1996sh,Strominger:1997eq}. The microstates of such black holes correspond to thermal excitations in a two-dimensional conformal field theory (CFT$_2$)  
and their entropy is counted by Cardy's formula \cite{Cardy:1986ie}, a universal CFT$_2$ formula that perfectly matches the area law on the gravity side. Moreover, the symmetries of the microscopic CFT  - two copies of the infinite-dimensional Virasoro algebra - are directly visible in spacetime in the form of  asymptotic symmetries \cite{Brown:1986nw}.

Unfortunately, no black hole in nature has an AdS$_3$ factor in its near-horizon region. Nevertheless,  a few years ago  it was proposed that  extremal Kerr black holes - examples of which do seem to exist in nature%
\footnote{Near-extremality ratios $|J|/M^2$ of 98\% have been claimed for the GRS 105+1915 \cite{Blum:2009ez} and Cygnus X-1 \cite{Gou:2013dna} black hole, as well as for the supermassive black holes in the active galactic nuclei MCG-6-30-15 \cite{Brenneman:2006hw} and 1H 0707-494 \cite{Fabian:2011ya}. Some independent measurements however point to non-extremal values, as reviewed in \cite{Fender:2010tk}.}
  -  are described by a (chiral half of a) two-dimensional  CFT \cite{Guica:2008mu}. This proposal was based on the study of the near-horizon scaling limit of the extremal Kerr black hole \cite{Bardeen:1999px} (henceforth abbreviated as NHEK), which enjoys an enlarged symmetry group, namely 
  $SL(2,\mathbb{R})_L \times U(1)_R$. The main supporting evidence was the enhancement of the $U(1)_R$ isometry to the full Virasoro algebra at the level of asymptotic symmetries and the perfect match between the black hole entropy formula and the Cardy entropy. These results were soon extended  (see the reviews \cite{Bredberg:2011hp,Compere:2012jk}) to very general extremal black holes, indicating the existence of a universal holographic correspondence for such black holes. The part of the geometry that appears to play the key role in the duality is a warped AdS$_3$ factor - whose structure is that of a $U(1)$  fibre over AdS$_2$ - which is universally present in the near-horizon region of extremal black holes \cite{Kunduri:2007vf}.

 Despite the remarkable agreement between the microscopic and macroscopic  entropy of extremal black holes, several puzzles remain, regarding the very applicability of Cardy's formula to the microscopic counting. First, Cardy's formula applies to two-sided CFT$_2$'s, whose symmetries consist of both a left-moving  and a right-moving Virasoro algebra. Nevertheless, all attempts to find an asymptotic symmetry group for NHEK (or, more generally, for warped AdS$_3$) that contains both copies of the Virasoro algebra \emph{simultaneously} have so far failed\footnote{Refs. \cite{Matsuo:2009pg,Rasmussen:2009ix,Castro:2009jf,Matsuo:2010ut,Azeyanagi:2011zj,Song:2011sr}  discuss boundary conditions that enhance  the $SL(2,\mathbb{R})_L$ factor to a left-moving Virasoro. }. Instead, boundary conditions for warped AdS$_3$ spacetimes have been found \cite{Compere:2008cv,Compere:2009zj}, which admit as asymptotic symmetries  a left-moving Virasoro algebra, together with a $U(1)$ Ka\v{c}-Moody algebra that enhances the right-moving translations. These results have inspired the search for two-dimensional QFTs - sometimes denoted as ``warped CFTs'' -  that exhibit these symmetries. While no consistent quantum example of a warped CFT is known to date  (see, however, the semi-classical examples in \cite{Compere:2013aya}), on very general grounds,  \cite{Hofman:2011zj} has shown that they are a natural extension of local two-dimensional QFTs with $SL(2,\mathbb{R})_L \times U(1)_R$ invariance,  and \cite{Detournay:2012pc} proved that  their density of states at high temperatures is dictated by a universal, Cardy-like formula%
\footnote{Note that if warped CFTs are relevant to understanding holography for warped $AdS_3$, as was suggested in \cite{Detournay:2012pc}, they cannot quite be the theories defined in \cite{Hofman:2011zj} because those were local, whereas it is expected on general grounds  \cite{Guica:2010sw} that the dual theories to warped AdS$_3$ would be non-local.}.

Second, 
as was pointed out in \cite{Amsel:2009ev,Dias:2009ex}, NHEK suffers from a problem common to all spacetimes that contain an AdS$_2$ throat, which is the ``no dynamics'' problem: any finite amount of energy in AdS$_2$ would destroy its asymptotics \cite{Maldacena:1998uz}; therefore, no bulk excitations of this spacetime are allowed. As a consequence, the only asymptotically NHEK spacetimes are NHEK itself and diffeomorphisms thereof, fact which is in tension with modular invariance and thus the applicability of Cardy's formula. In particular, there are no asymptotically NHEK spacetimes that could correspond to the vacuum or to states  whose right-moving temperature $T_R \neq \frac{1}{2\pi}$\footnote{Another puzzle concerns the fact that Cardy's formula is applied in the $\mathcal{O}(1)$-temperature regime, which lies outside its usual range of validity. This is, nevertheless, a very common feature of AdS$_3$ black holes and \cite{Hartman:2014oaa} has recently given a  beautiful explanation for why the applicability of Cardy's formula is extended for CFTs with a large central charge and a sparse light spectrum. We might expect that similar arguments also apply to Kerr/CFT. For an attempt to use string dualities  to bring the extreme Kerr black hole in a regime where Cardy's formula is valid, see \cite{Jejjala:2009if}.}.

Finally, the NHEK throat admits so-called travelling wave perturbations \cite{Bardeen:1999px}, which have an oscillatory behaviour at infinity and can carry energy and momentum through the spacetime boundary. The stability or instability of NHEK under such perturbations is a subtle question of boundary conditions: while \cite{Amsel:2009ev}, who required that no flux pass through the boundary of the spacetime, found an exponential growth of the perturbations with time, \cite{Dias:2009ex}, who imposed purely outgoing boundary conditions instead, found an exponential damping, as  \cite{Bardeen:1999px} had predicted
. Thus, one seems to have a choice between the symplectic form not being conserved and instabilities in the near-horizon region. In either case, one cannot define a self-consistent phase space, even at the linear level.

Some of the above problems do have a simple resolution if we consider the Kerr/CFT correspondence embedded in its larger context as holography for warped AdS$_3$ spacetimes, relation which was made precise in \cite{ElShowk:2011cm}. For example, the no dynamics puzzle has a similar origin and resolution to its AdS$_3$/CFT$_2$ counterpart - for a clear explanation, see \cite{Balasubramanian:2009bg}. The lack of dynamics  is due to taking an infrared  limit (also known as the ``very near-horizon'' limit \cite{Strominger:1998yg}) that only keeps the ground states of the system. In order to avoid it, one should seek a different decoupling limit \cite{Maldacena:1997re,Maldacena:1998bw}, which also keeps  some finite energy excitations from the point of view of the decoupled theory. While this intermediate scaling limit is not known  (to exist) 
for Kerr
, it can be found for many type IIB geometries,  via an analysis identical to that in \cite{ElShowk:2011cm}. These decoupled geometries, many examples of which have been studied in \cite{sss}, have full dynamics and 
 contain a rich spectrum of warped AdS$_3$ black strings, whose thermodynamics is identical to that of the  BTZ black string. In certain cases, the spectrum of black string solutions can
   be entirely captured within a three-dimensional consistent truncation of type IIB supergravity  \cite{sss}. One of these truncations, which will be studied in this article, is particularly interesting because its black string solutions, upon dimensional reduction and taking the very near-horizon limit, can model the near-horizon geometry of the charged generalization of the singly-spinning, extremal five-dimensional Myers-Perry black hole \cite{ElShowk:2011cm} (see also \cite{Guica:2010ej,Dias:2007nj}).

 The main goal of the present paper is to address the first puzzle that we mentioned, namely the construction of a second copy of the Virasoro symmetry in warped AdS$_3$, which acts \emph{simultaneously} with the first. In order to have a full-fledged, dynamical, holographic correspondence, we will be working in the context of the type IIB consistent truncations mentioned above. The presence of both Virasoro symmetries in this  context has been long forecast by the holographic analyses of \cite{fg,decrypt}, who showed that the holographic stress tensor for the warped AdS$_3$ spacetime is at the same time symmetric, conserved, and traceless, and is thus expected to generate two copies of the Virasoro algebra. In this paper, we show this is indeed the case (for a given set of ``Dirichlet'' boundary conditions), using the more transparent covariant phase space formalism. Concretely, we show that for each consistent choice of boundary conditions in an auxiliary AdS$_3$ space-time, there exists a choice of boundary conditions in warped AdS$_3$, such that the associated conserved charges are identical. 
%
We can thus easily obtain both Virasoro $\times$ Virasoro and Virasoro $\times$ Ka\v{c}-Moody asymptotic symmetry groups, associated respectively to Dirichlet (Brown-Henneaux) \cite{Brown:1986nw} and mixed chiral boundary conditions \cite{Compere:2013bya} in the auxiliary AdS$_3$.


The statement above applies at full non-linear level, but only to a certain,  non-dynamical sector of the theory. The most interesting question nevertheless is whether the Virasoro $\times$ Virasoro symmetry extends to the full phase space of the theory, which includes bulk propagating degrees of freedom. If the answer is yes, then we would have shown that boundary conditions exist such that warped AdS$_3$ is holographically dual to a two-dimensional CFT (at least in the semi-classical limit), in the sense that all excitations that have a gravity dual transform under representations of the two Virasoro algebras, upon canonical quantization of the supergravity fields\footnote{It is important to note however that this dual theory cannot be a usual CFT$_2$, since the background only has  $SL(2,\mathbb R) \times U(1)$ invariance and the dual theory is non-local, as pointed out by the structure of counterterms in holographic renormalization \cite{Guica:2010sw} and by concrete examples in string theory \cite{Bergman:2001rw
}.}. We are indeed able to show that these symmetries extend to the entire linearized phase space of gravity in warped AdS$_3$, as long as no travelling waves are present. We  leave the analysis  of the conformal symmetries in  presence of the travelling waves to future work.

This paper is organised as follows. In section \ref{setup}, we review two of the consistent truncations of \cite{sss} that we will work with, as well as the covariant phase space formalism. In section \ref{sec:bnd}, we study  slices through the phase space of these theories that do not contain bulk propagating modes, and show that their symplectic structure is isomorphic to that of pure three-dimensional Einstein gravity with a negative cosmological constant. This proves our claim that there is a one-to-one map between the conserved charges in AdS$_3$ and warped AdS$_3$; we work out two explicit examples  in sections \ref{excon} and \ref{dnbc}. In section \ref{propmodes}, we discuss the spectrum of linear perturbations in the theories of interest and show that, even  in presence of the propagating modes, the symplectic form can always be made finite. We gather some details of the computations in the appendices.

\section{Setup and review \label{setup}}

We begin this section by reviewing two consistent truncations\footnote{For other consistent truncations of string/M-theory that contain warped AdS$_3$ solutions, see  \cite{Colgain:2010rg,Karndumri:2013dca}. } of type IIB supergravity theory to three dimensions that were worked out in \cite{sss} and that will be the main examples we  use in this article. These truncations are interesting because they contain a rich spectrum of warped analogues of the BTZ black string as solutions, i.e. solutions characterized by two independent conserved charges that can be interpreted as excited states above the warped AdS$_3$ vacuum\footnote{Given how easy it is to construct any given warped AdS$_3$ spacetime using a (topologically) massive vector (i.e. using a combination of Yang-Mills/Chern-Simons terms or, equivalently \cite{Deser:1984kw}, Proca/Chern-Simons terms \cite{Banados:2005da}), it is quite surprising how difficult it is to find models that contain black hole/string solutions with both of the above properties. }. 
%
%
%
%
These warped black strings  can be obtained by applying certain solution-generating techniques to the D1-D5 solution of type IIB string theory. For details on how the solutions were obtained, the consistent truncation Ans\"{a}tze, the terminology and the possible microscopic brane interpretation of these backgrounds, please consult \cite{sss,Bena:2012wc}.

In the last two subsections, we review the covariant phase space formalism  \cite{Lee:1990nz,Iyer:1994ys} for the construction of the symplectic form and conserved charges in diffeomorphism-invariant theories.

\subsection{The S-dual dipole truncation \label{sddip}}

The simplest known consistent truncation of string theory that admits warped AdS$_3$ solutions is the so-called S-dual dipole truncation \cite{sss}, with action given by 
\be
S= \frac{1}{16 \pi G_3} \int d^3 x \sqrt{g} \left( R - 4 (\p U)^2 - \frac{4}{\ell^2} \, e^{-4U} \,  A^2 + \frac{2}{\ell^2} \, e^{-4U} (2-e^{-4U}) -\frac{1}{\ell} \, \e^{\mu\nu\rho} A_\mu F_{\nu\rho}\right) .\label{actsddip}
\ee
This theory describes  one scalar field $U$ and one chiral massive vector  $A_\mu$, 
for a total of two local degrees of freedom. The corresponding equations of motion are
\be
F_{\mu\nu} = \frac{2}{\ell}\, e^{-4U} \e_{\mu\nu\l} A^\l,\qquad \frac{\ell^2}{2} e^{4U}\, \Box U = 1 - e^{-4U} -A^2 ,\label{eomA}
\ee
\be
R_{\mu\nu} + \frac{2}{\ell^2} \, e^{-4U} (2-e^{-4U}) \, g_{\mu\nu} = 4 \p_\mu U \p_\nu U + \frac{4}{\ell^2} \, e^{-4U} A_\mu A_\nu.
\ee
In the first half of this article, we will be interested in solutions to the above equations that have constant $U$. For $U=0$, this theory admits a null warped AdS$_3$ solution, given by
\be
ds^2 = \ell^2 \left( -\l^2 r^2 du^2 + 2 r\, du dv + \frac{dr^2}{4r^2} \right) \label{nwmet}
\ee
and
\be
A = \l \ell r du ,
\ee
where $\l$ is an arbitrary parameter. For $U$ positive, the theory admits a family of spacelike warped black string solutions, parametrized by the left/right moving temperatures\footnote{The parameters $T_\pm$ are related to the usual right/left-moving temperatures by $T_\pm = \pi T_{R,L}$. Also, in our notation,  the direction $v$ is referred to as right-moving ($+$), and $u$ as left-moving $(-)$.} $T_\mp$ and $\lambda$
\be
ds_3^2 = \ell^2 \left(T_+^2\, dv^2 + 2 r \, du \, dv + \left[\, T_-^2\, (1+  \l^2 T_+^2) -  \l^2 r^2\right] du^2 + \frac{ (1+  \l^2 T_+^2) \,  dr^2}{4 (r^2 - T_+^2 T_-^2)} \right),\label{bsmet}
\ee
\be
e^{4U} = 1+\l^2 T_+^2,\qquad \quad A= \frac{\l \ell}{\sqrt{1+ \l^2 T_+^2}} (T_+^2 dv + r du) \label{Abh},
\ee
whose thermodynamic properties are identical to those of the BTZ black string \cite{sss}. It is useful to note that $A^\mu$ is a Killing vector of the metric \eqref{bsmet}, $A^\mu = \l \ell^{-1} (1+\l^2 T_+^2)^{-\half} \,\p_v$. 

Unlike the BTZ black strings, which are diffeomorphic to Poincar\'e AdS$_3$, here only the $T_-$ dependence of the metric \eqref{bsmet} can be turned off by a coordinate transformation. There is no diffeomorphism that relates  black strings of different $T_+$, as can be noted from the fact that the solutions have different values of the scalar $U$. 
 
The  Poincar\'e vacuum \eqref{nwmet} can be simply obtained by setting $T_+=T_- =0$. By analytic continuation, the global null warped background of \cite{Blau:2009gd} corresponds to $T_+=0$, $T_-=i$.
 Solutions with negative $U$ can be obtained from the analytic continuation $T_+^2 \rightarrow -T_+^2$, but the parameters must be restricted to $\l T_+ < 1$. These  solutions, which represent timelike warped AdS$_3$ space-times, are not very well understood (see however \cite{Compere:2009zj}). 


\subsection{The ``NHEMP'' truncation \label{nhek}}
Another consistent truncation worked out in \cite{sss} is what we call the ``NHEMP'' truncation\footnote{The same truncation was called ``NHEK'' in \cite{sss}. Here we use instead the acronym NHEMP, for greater accuracy.}, which models the near-horizon limit of the six-dimensional uplift of the  singly-rotating, five-dimensional charged generalization of the Myers-Perry black hole \cite{Cvetic:1996xz,Dias:2007nj}, reduced to three dimensions. 
The truncated three-dimensional action contains two scalars $U_{1,2}$ and two  chiral massive vector fields $A_{1,2}$, for a total of four propagating degrees of freedom\footnote{In terms of the notation in \cite{sss}, we have $U_1=U,\ U_2 = V,\,A_1 = A,\,A_2 = \hat A$ and $\l_{here}=\tilde \l_{there}$.}
\bea
S &=&  \frac{1}{16 \pi G_3} \int d^3x \sqrt{-g} \left[ R - 6 (\p U_1)^2 -2(\p U_2)^2 - 4 \p U_1 \cdot \p U_2  - \frac{ 4}{\ell^2} \,  e^{-4U_1} (A_1 -A_2)^2 - \right. \non \\
&& \hspace{-1 cm} \left.  -\frac{ 8}{\ell^2} \, e^{-4U_1-4U_2}  A_2^2+ \frac{2}{\ell^2} \, e^{-8U_1}  \left(4\, e^{2U_1-2U_2} -1 - 2 e^{-4U_2} \right) -\frac{2}{\ell}\, \e_{\mu\nu\rho} \left( A_1^\mu -\half  A_2^\mu \right)  F_2^{\nu\rho}\right] .\label{nhekact}
\eea
The associated equations of motion read
\bea
{}^{(3)}R_{\mu\nu}&=& 6\, \p_\mu U_1 \p_\nu U_1 + 2\, \p_\mu U_2 \p_\nu U_2 + 2 \,(\p_\mu U_1 \p_\nu U_2 + \p_\nu U_1 \p_\mu U_2)+ \frac{8}{\ell^2} \,e^{-4U_1-4U_2} \,  A_{2 \mu}  A_{2\nu} + \non \\
& + & \frac{4}{\ell^2}\, e^{-4U_1} \, ( A_{1 \mu} -A_{2\mu})( A_{1\nu}-A_{2\nu}) - \frac{2}{\ell^2} \, g_{\mu\nu} \,e^{-8U_1} \left(4\, e^{2U_1-2U_2} -1 -2 e^{-4U_2}  \right) ,\label{eq1}
\eea

\be
e^{4U_1+4U_2} \star_3 F_1= - \frac{4}{\ell}  A_2\;, \;\;\;\;\;   e^{4U_1} \star_3  F_2 = - \frac{2}{\ell} (  A_1 -  A_2),\label{eq2}
\ee

\be
\Box U_1 = \frac{2}{\ell^2} \, e^{-8U_1} (2 e^{2U_1-2U_2} - 1 - e^{-4U_2}) - \frac{2}{\ell^2} \, e^{-4U_1} \left( A_1 - A_2\right)^2,\label{eq3}
\ee

\be
\Box U_2 = \frac{2}{\ell^2} \, e^{-8U_1} (1- e^{-4U_2}) -\frac{8}{\ell^2}\, e^{-4U_1 -4U_2}  A_2^2 + \frac{2}{\ell^2} \, e^{-4U_1} \left( A_1 - A_2\right)^2.\label{eq4}
\ee
The above action admits warped black string solution  parametrized by the deformation parameter $\lambda$ and the left/right moving temperatures $T_\pm$. The metric is the same as before
\be
ds_3^2 =\ell^2 \left( T_+^2 d v^2 + 2  r d u dv + [T_-^2 (1+  \l^2 T_+^2) -  \l^2  r^2] du^2  +\frac{ d r^2}{ 4( r^2 - T_+^2 T_-^2)} (1+  \l^2 T_+^2) \right). \non
\ee
The two chiral massive vector fields  read
\be
A_{1,2} =\a_{1,2}\, A\;, \qquad A \equiv \frac{ \lambda \, \ell}{\sqrt{1+ \l^2 T_+^2}}\, (rdu + T_+^2 dv) ,
\ee
where $\a_{1,2}$ are rather complicated functions of $\l T_+$, as are the two scalar fields $U_{1,2}$. The solution for $U_{1}$ is
\be
e^{4 U_1} = \frac{1}{4}\, \left(1+ \sqrt{9+8 \l^2 T_+^2}\right)
\ee
and the other three constants are given by
\be
e^{2U_2} = \frac{1+\l^2 T_+^2}{ e^{6 U_1} }\;, \;\;\;\;\; \a_1 = \frac{2 e^{4U_1}}{\sqrt{4 e^{8U_1}-1}} \;, \;\;\;\;\;\a_2 = \sqrt{\frac{2 e^{4U_1}-1}{2e^{4U_1}+1}} .\label{solsa}
\ee
%
%
%
For $T_\pm=0$, we recover the corresponding null warped AdS$_3$ solution, whereas for $T_+^2 <0$ (within reasonable ranges), we obtain asymptotically timelike warped AdS$_3$ spacetimes.

\subsection{Review of the covariant phase space formalism}
\label{symp}

The main goal of this article is to study the structure of the phase space of theories admitting  warped AdS$_3$ solutions, such as the two theories we just described.
%
%
We thus review some general results on the construction of the symplectic structure in gravity, including the explicit expressions for the theories at hand. 

\subsection*{Generalities}

Let us first introduce some definitions. Let the fields in the theory (including the metric) be collectively denoted as $\phi = \{ \phi^i \}$, and let the Lagrangian $n$-form be denoted by $\boldsymbol{ \mathcal{L}}[\phi]$. We define the presymplectic $n-1$ form $\boldsymbol \Theta [\phi,\d \phi ]$ via
\be
\d \boldsymbol{\mathcal{L}} [\phi]= \boldsymbol E_{\phi^i}[\phi] \d \phi^i + \boldsymbol d \boldsymbol \Theta [\phi, \d \phi]
\ee 
where $\boldsymbol E_{\phi^i}[\phi]$ are the Euler-Lagrange equations for the field $\phi^i$. The symplectic $n-1$ form $ \boldsymbol \om [\phi, \d_1 \phi, \d_2 \phi ]$ is defined as
\bea
{\boldsymbol \omega}[ \phi,\delta_1 \phi ,\delta_2 \phi] = \delta_1 {\boldsymbol  \Theta}[\phi, \delta_2 \phi] - \delta_2 {\boldsymbol \Theta}[\phi,\delta_1 \phi] .\label{sympl}
\eea
From the start, we note that the presymplectic form suffers from at least two ambiguities. First, one can change the action by adding a boundary term $\boldsymbol{ \mathcal{L}} \rightarrow \boldsymbol{\mathcal{L}} + \boldsymbol d \boldsymbol \mu$, which changes the presymplectic form as $\boldsymbol \Theta \rightarrow \boldsymbol \Theta + \delta \boldsymbol \mu[\phi]$. However, this ambiguity cancels in the symplectic structure and therefore has no physical significance in the construction of a phase space. 

Second, the presymplectic form $\boldsymbol \Theta$ is only defined up to the addition of an exact form 
\be
\boldsymbol \Theta \rightarrow \boldsymbol \Theta + \boldsymbol d \boldsymbol Y[\phi,\delta \phi] .\label{ambig}
\ee
Under this replacement, the symplectic structure is shifted by a boundary term
\bea
{\boldsymbol \omega}[ \phi,\delta_1 \phi ,\delta_2 \phi] \r {\boldsymbol \omega}[ \phi,\delta_1 \phi ,\delta_2 \phi] + \boldsymbol d {\boldsymbol \omega}_{\boldsymbol Y}[ \phi,\delta_1 \phi ,\delta_2 \phi] \label{ambiom}
\eea
where
\be
{\boldsymbol \omega}_{\boldsymbol Y}[ \phi,\delta_1 \phi ,\delta_2 \phi] = \delta_1 {\boldsymbol  Y}[\phi, \delta_2 \phi] - \delta_2 {\boldsymbol  Y}[\phi,\delta_1 \phi] \label{omY}
\ee
This boundary term will turn out to play a crucial role in our analysis. We can define a representative $\boldsymbol \Theta$ by a standard algorithm, which consists in integrating  by parts the variation of the Lagrangian or, more formally, by acting on the Lagrangian with Anderson's homotopy operator $\boldsymbol I^n_{\delta \phi}$  \cite{Andersonbook,Barnich:2001jy,Barnich:2007bf}, defined for second order theories as
\bea
\boldsymbol \Theta^{ref}=\boldsymbol I^n_{\delta \phi}  \boldsymbol{\mathcal{L}} \;,\;\; \;\;\;\;\; \boldsymbol I^n_{\delta \phi}  \equiv \left( \delta \phi^i \frac{\p}{\p \p_\mu \phi^i} - \delta \phi^i \p_\nu \frac{\p}{\p \p_\nu \p_\mu \phi^i} \right) \frac{\p}{\p \boldsymbol d x^\mu} .\label{homot}
\eea
The total presymplectic structure is therefore $\boldsymbol \Theta = \boldsymbol \Theta^{ref} +\boldsymbol d \boldsymbol Y $, where $\boldsymbol Y[\phi,\delta \phi] $ still needs to be specified. 

Finally, it is possible to give an independent definition of the symplectic structure, which  differs from \eqref{sympl} by a boundary term\footnote{This is the so-called invariant symplectic structure, defined in \cite{Barnich:2007bf} as $\boldsymbol W = \frac{1}{2}\boldsymbol I^n_{\delta_1 \phi} (\boldsymbol E_{\phi^i} \delta_2 \phi) - (1 \leftrightarrow 2) = \boldsymbol \omega + \boldsymbol d \boldsymbol E$, where $\boldsymbol E[\phi,\delta_1 \phi,\delta_2 \phi]$ is a non-$\delta$ exact boundary term.}. In this paper we shall not consider the latter ambiguity. 


\noindent Next, given an arbitrary vector $\xi$, the Noether current
\be
\boldsymbol J_\xi =  \boldsymbol \Theta [\phi,\d_\xi \phi] - \xi \cdot  \boldsymbol{\mathcal{L}}[\phi]
\ee
is closed on-shell\footnote{The proof uses the identity%
$$\d_\xi \boldsymbol \Lambda = \xi \cdot \boldsymbol d \boldsymbol \Lambda + \boldsymbol d (\xi \cdot \boldsymbol \Lambda)$$ \vspace{-4mm}} and therefore exact, $\boldsymbol J_\xi = \boldsymbol d \boldsymbol Q_\xi[\phi]$, as a consequence of the absence of non-trivial cohomology \cite{Lee:1990nz,Andersonbook,Barnich:2000zw}. It is easy to show that, on-shell
\be
\d \boldsymbol  J_\xi =\boldsymbol  \om
[\phi,\d \phi, \d_\xi \phi] + \boldsymbol  d (\xi \cdot \boldsymbol  \Theta [\phi,\d \phi]) = \boldsymbol d \d \boldsymbol  Q_\xi[\phi]
\ee
from which there follows that (still on-shell)
\be
\boldsymbol  \om
[\phi, \d \phi, \d_\xi \phi] = \boldsymbol d \boldsymbol k_\xi [\delta \phi,\phi] 
\label{chom1}
\ee
where $\boldsymbol k_\xi [\delta \phi,\phi] $ is given by
\be
\boldsymbol k_\xi [\delta \phi,\phi] \equiv \d \boldsymbol  Q_\xi [\phi] - \xi \cdot \boldsymbol  \Theta [\phi,\d\phi] .
\label{consch}
\ee
When $\xi$ is a \emph{symmetry}, $\d_\xi \phi =0$, the left-hand side of \eqref{chom1} is zero and $\boldsymbol k_\xi [\delta \phi,\phi] $
is nothing but the infinitesimal conserved charge associated with $\xi$. Consistently with theorems in cohomology \cite{Barnich:2000zw, Barnich:2001jy}, the boundary contribution $\boldsymbol  \om_{\boldsymbol Y} $ vanishes identically  for exact symmetries and the conserved charge is  unaffected by the ambiguity \eqref{ambig}.

The situation is rather different  when $\xi$ is an \emph{asymptotic symmetry} of a given set of boundary conditions. The boundary conditions must be chosen such that the total symplectic form 
 is asymptotically zero for arbitrary perturbations. More precisely, we require that\footnote{Here $o(r^{-c})$ is any function which, when multiplied by $r^{c}$, asymptotes to zero in the limit $r\rightarrow \infty$.}
\bea
\boldsymbol \om_{ab} [\phi, \d_1 \phi, \d_2 \phi] = o(r^0), \qquad\boldsymbol \om_{r a} [\phi, \d_1 \phi, \d_2 \phi] = o(r^{-1}),\label{phs}
\eea
where $a$ denotes the tangent indices to the boundary, itself  located at $r \rightarrow \infty$. The first equation above represents the requirement that the symplectic flux be conserved, whereas the second corresponds to normalizability of the symplectic form at infinity. 
 Note that we only require that $\boldsymbol \om$ satisfy the above boundary conditions for some choice of $\boldsymbol \Theta$ in \eqref{ambig} or, equivalently, having fixed a reference $\boldsymbol \Theta$ as e.g. in \eqref{homot}, only for some particular  choice of boundary term $\boldsymbol Y$. The contribution of this boundary term to the asymptotically conserved charge is 
\be
\boldsymbol k_\xi [\delta \phi,\phi] = \boldsymbol k_\xi^{ref} [\delta \phi,\phi] + \d  \boldsymbol Y[\phi , \delta_\xi \phi] - \xi \cdot \boldsymbol d \boldsymbol Y [\phi, \delta \phi].
\ee
where $\boldsymbol k_\xi^{ref}$ is computed using $\boldsymbol \Theta = \boldsymbol \Theta^{ref}$. When \eqref{phs} is satisfied, it follows from \eqref{chom1} that the charge \eqref{consch}
is finite and asymptotically conserved. 
Assuming $\xi^r =\mathcal{O}(r),\, \xi^a = \mathcal{O}(r^0)$, the condition \eqref{phs} is sufficient to ensure integrability, since
\bea
\delta_1 \boldsymbol k_\xi [\delta_2 \phi,\phi]  - (1 \leftrightarrow 2)= - \xi \cdot \boldsymbol \omega[\phi, \delta_1 \phi,\delta_2 \phi] = o(r^0).
\eea
To summarize, a phase space exists if and only if there exists a boundary symplectic structure such that \eqref{phs} can be enforced for any element $\phi$ in the phase space and any linear perturbations $\d_1 \phi$, $\d_2 \phi$ around it. 



\subsection{Explicit expressions for the symplectic structure and charges}

The two consistent truncations presented in sections \ref{sddip} and \ref{nhek} can be summarized via an action of the form
\be
S = \frac{1}{16 \pi G_3} \int d^3 x \sqrt{-g} \Big( R -\half\, f_{ab} \, \p_\mu U^a \p^\mu U^b -
 \half \, b_{ij} A^i_\mu A^{j\,\mu}   + \half \, c_{ij}\,\e^{\mu\nu\rho} A^i_\mu F^j_{\nu\rho}   \Big)  \label{act3d1}
\ee
where all coefficients can in principle be functions of the scalar fields $U^a$. 
For the S-dual dipole theory, we have
\be
f = 8 , \quad \quad b= \frac{8}{\ell^2}\, e^{-4U}  , \quad \quad c = - \frac{2}{\ell} ,
\ee
whereas for the NHEMP truncation
\be
f=  \left( \begin{array}{cc} 12 & \; 4 \\ 4 & \;4 \end{array} \right),\quad \quad 
b = \frac{8\, e^{-4U_1}}{\ell^2} \left( \begin{array}{cc} 1 & -1 \\ -1 & 1+2 e^{-4U_2} \end{array} \right),\quad \quad
c = \frac{1}{\ell} \left( \begin{array}{cc} 0 & -4 \\ 0 &\; 2 \end{array} \right). \label{nhekdata}
\ee
For theories with action of the above form, the pre-symplectic form is given by 
\be
\boldsymbol \Theta^{ref} = \frac{1}{16 \pi G_3}  \left( \boldsymbol \Theta_g+ \boldsymbol \Theta_{CS} + \boldsymbol \Theta_{scal} \right). \label{thsplit}
\ee
In three spacetime dimensions, it is oftentimes useful to work in terms of the Hodge duals
\be
\Theta^\mu =  -\half \, \e^{\mu\nu\rho} \Theta_{\nu\rho} \;, \;\;\;\;\;  \om^\mu =  -\half \, \e^{\mu\nu\rho} \om_{\nu\rho} \label{Hodge}
\ee 
where $\e$ is the Levi-Civita tensor density 
We then have  ${\boldsymbol \Theta} = \frac{1}{2}\, \Theta^\mu  \e_{\mu\alpha\beta}\, dx^\alpha \wedge dx^\beta$ and $\Theta^\mu$ transforms as a vector. Using this notation,
 the Einstein-Hilbert, Chern-Simons and scalar contributions to the presymplectic form are given by
\be \label{psf}
\Theta^\mu_g=  \nabla_\l h^{\l \mu} - \nabla^\mu h \;, \;\;\;\;\; \Theta^\mu_{CS} = c_{ij} \e^{\mu\nu\rho} A_\rho^i  \d A_\nu^j \;, \;\;\;\;\; \Theta^\mu_{scal} = - f_{ab} \nabla^\mu U^b \d U^a 
\ee
where $h_{\mu\nu} = \d g_{\mu\nu}$. As discussed in the previous subsection, $\boldsymbol \Theta$ is ambiguous up to the addition of the boundary term \eqref{ambig}. The corresponding bulk symplectic form reads
\be
\boldsymbol\om
= \frac{1}{16 \pi G_3}\, (\boldsymbol\om_g 
 + \boldsymbol\om_{CS} + \boldsymbol\om_{scal}).\label{wisym}
\ee
The Einstein contribution is 
\bea
\om^\mu_g[g,\delta_1 g,\delta_2 g] &=&\frac{1}{2}P_{LW}^{\mu\alpha\beta\gamma\delta\epsilon} \left(h_{2 \alpha\beta} \nabla_\gamma h_{1\delta \epsilon} -   h_{1 \alpha\beta} \nabla_\gamma h_{2\delta \epsilon} \right), \\
&=&  \frac{1}{2} \left[ ( 2\nabla_\l h_1^\mu{}_\rho - \nabla^\mu h^1_{\l\rho}) h_2^{\l \rho} - \nabla_\rho h_1 \, h_2^{\rho \mu} + h_1 (\nabla_\rho h^{\rho\mu}_2 - \nabla^\mu h_2) - (1 \leftrightarrow 2) \right] \nn
\eea
where $h_{1\mu\nu} = \delta_1 g_{\mu\nu}$, $h_{2\mu\nu} = \delta_2 g_{\mu\nu}$ and the Lee-Wald symbol \cite{Lee:1990nz} is
\bea
P_{LW}^{abcdef}&=&2 g^{ae}g^{fb}g^{cd}- g^{ad}g^{be}g^{fc}-g^{ab}g^{cd}g^{ef}-g^{bc}g^{ae}g^{fd}+g^{bc}g^{ad}g^{ef}.
\eea
The Chern-Simons and scalar contributions are
%
\be
\om^\mu_{CS} = \e^{\mu\a\b} \d_1 ( c_{ij} A^i_\b ) \, \d_2 A^j_\a - (1 \leftrightarrow 2),
\ee
\be
\om^\mu_{scal} = - \d_1 (f_{ab} \nabla^\mu U^b) \, \d_2 U^a - \half \, h_1 f_{ab} \nabla^\mu U^b \, \d_2 U^a - (1 \leftrightarrow 2).
\ee
%
%
In taking the variations, one needs to be careful about the possible dependence of $c_{ij}$ and $f_{ab}$ on the scalars $U^a$.

The infinitesimal conserved charges are entirely determined by the symplectic form, as one can see from \eqref{chom1}.  
 For the action \eqref{act3d1}, it has been shown in \cite{Compere:2009dp} that the conserved charge $\boldsymbol k_\xi[\phi , \delta \phi]$ associated to $\xi^\mu$ has the following expression
\be
\boldsymbol k^{cov}_\xi[\phi , \delta \phi] = \half \, \e_{\mu\nu\l} K_\xi^{\mu\nu}[\phi,\delta \phi] \, dx^\l \label{ecc}
\ee
with
\be
 K^{\mu\nu}_\xi = \frac{1}{8\pi G_3} \left( K^{\mu\nu}_g 
+ K^{\mu\nu}_{CS} + K^{\mu\nu}_{scal} \right).
\ee
The three contributions are
\be
K^{\mu\nu}_g= \xi^\nu \nabla^\mu h - \xi^\nu \nabla_\s h^{\mu\s} + \xi_\s \nabla^\nu h^{\mu\s} + \half h \nabla^\nu \xi^\mu - h^{\rho \nu} \nabla_\rho \xi^\mu ,
\ee
\be
K^{\mu\nu}_{CS} =  c_{ij} \e^{\mu\nu\rho} \d A_\rho^j \, A_\l^i \xi^\l, \qquad 
K^{\mu\nu}_{scal}=  \xi^\nu f_{ab} \nabla^\mu U^b \d U^a.
\ee

\section{Phase spaces without bulk propagating modes}
\label{sec:bnd}

It is well-known that pure gravity in three dimensions has no bulk propagating degrees of freedom. In particular, all solutions of three-dimensional  Einstein gravity with a negative cosmological constant are locally diffeomorphic to AdS$_3$. Despite its apparent triviality, the theory contains a rich black hole spectrum \cite{Banados:1992wn,Banados:1992gq}, solutions with non-trivial topology  \cite{Aminneborg:1997pz,Skenderis:2009ju} and allows for non-trivial boundary excitations known as boundary gravitons \cite{Brown:1986nw} or photons \cite{Compere:2013bya}. The latter are obtained by exponentiating the generators of ``large'' diffeomorphisms: diffeomorphisms  that fall off slowly  enough at infinity to give rise to non-trivial conserved charges. These modes are very interesting from a holographic point of view, as they represent excitations of  the stress tensor of the holographically  dual theory \cite{Strominger:1997eq}. 

The three-dimensional theories of gravity coupled to matter that we discussed in section \ref{setup} admit an equally rich subsector of degrees of freedom that do not propagate into the bulk. This subsector can be obtained by restricting the phase space of these theories to a slice  on which the scalar fields are constant.  As it was shown in \cite{fg, decrypt}, the solution space of the original theories, when restricted to this slice, is the same as the solution space of pure three-dimensional Einstein gravity with a negative cosmological constant. Thus, the solutions on the slice  can be entirely characterized by  an auxiliary AdS$_3$ metric $\hat g_{\mu\nu}$, which was argued to encode the holographic stress tensor data for warped AdS$_3$ \cite{fg,decrypt}. 

In the following, we review the construction of this non-bulk-propagating subsector of the consistent truncations and we show that not only is the solution space the same as that of three-dimensional Einstein gravity, but also the symplectic structures of the two theories can be made identical.  In turn, this implies that there is a one-to-one map between conserved charges in AdS$_3$ and warped AdS$_3$; in particular, the asymptotic symmetry groups are the same. 
%
We exemplify this construction of a consistent  phase space for warped AdS$_3$ in sections \ref{excon} and \ref{dnbc}, using two known sets of boundary conditions for the auxiliary AdS$_3$, namely Dirichlet  \cite{Brown:1986nw} and mixed Dirichlet-Neumann chiral boundary conditions \cite{Compere:2013bya}.

\subsection{Universal solution space without bulk propagating modes}
\label{univ}

We are interested in solutions to the equations of motion which are not dynamical in the bulk. As can be noted from section \ref{propmodes}, all linearized bulk propagating modes have non-trivial scalar profiles, which are moreover linearly independent. Since we are only interested in solutions obtained by perturbative expansion around the black string backgrounds, it is therefore sufficient to set the scalar fields to constants 
in order to restrict to the non-bulk-propagating ones. Below we  review the explicit construction of these solutions  for both the S-dual dipole and NHEMP truncations, following \cite{decrypt}.

\subsubsection*{The S-dual dipole truncation}


Since setting $U$ constant turns off  the bulk propagating modes, all  the remaining solutions to the equations of motion must be diffeomorphic to the black string backgrounds \eqref{bsmet} - \eqref{Abh}. The  most general background that is diffeomorphic to the black strings can be found by rewriting \eqref{bsmet}-\eqref{Abh} using a completely covariant notation, for some fixed $U$. 
 We find that 
\be
g_{\mu\nu} = e^{4U} (\hat g_{\mu\nu} - A_\mu A_\nu )\label{defgh}
\ee
where $\hat g_{\mu\nu}$ satisfies the pure Einstein equations of motion in three dimensions
\be
\hat R_{\mu\nu} +\frac{2}{\ell^2} \,\hat g_{\mu\nu} = 0. \label{eqhatg}
\ee
The vector field $A_\mu$ satisfies its usual equations of motion \eqref{eomA}, restricted to $ U=$ constant,
\be
F_{\mu\nu} = \frac{2}{\ell} \, e^{-4U} \e_{\mu\nu\l} A^\l \;, \;\;\;\;\; A^2 = 1-e^{-4U}.\label{kv}
\ee
It is nevertheless useful to rewrite the above equations in terms of $\hat g_{\mu\nu}$, using  \eqref{defgh} 
\be
F_{\mu\nu} = \frac{2}{\ell}\,\hat \epsilon_{\mu\nu\lambda}\hat A^\lambda,\qquad \hat A^2= \hat  A^\mu A_\mu = 1-e^{-4U} \;, \;\;\;\;\; \hat A^\mu \equiv \hat g^{\mu\nu} A_\nu .\label{consAb}
\ee
Finally, one notices that $A_\mu$ is a Killing vector\footnote{It is an interesting question whether one can  derive \eqref{killv}  directly from \eqref{consAb}. A proof may be possible by  noting that $\hat{\mathcal{K}}_{\mu\nu} \equiv \hat \nabla_\mu A_\nu + \hat \nabla_\nu A_\mu$ satisfies
\be
\hat{\mathcal{K}}_{\mu\nu} \hat A^\nu =0 \;, \;\;\;\;\; \hat{\mathcal{K}}^\mu{}_\mu =0\;, \;\;\;\;\; \hat \nabla^\mu \hat{\mathcal{K}}_{\mu\nu} =0 
\ee 
\vspace{-5mm}
}  of both  $g_{\mu\nu}$ and $\hat g_{\mu\nu}$
\be
 \nabla_\mu A_\nu +  \nabla_\nu A_\mu = \hat \nabla_\mu A_\nu + \hat \nabla_\nu A_\mu =0. \label{killv}
\ee
%
Equation \eqref{eqhatg} implies that $\hat g_{\mu\nu}$ is locally $AdS_3$. A general solution to this equation can be constructed via e.g. the Fefferman-Graham expansion. 
The claim is that,  given a solution for $\hat g_{\mu\nu}$, one can always construct  a solution for $A_\mu$ - and thus $g_{\mu\nu}$ - that does not contain any propagating modes. 

There are two ways of solving for  $A_\mu$. One can start by solving the first equation in \eqref{consAb}, which defines  self-dual vector field $A_\mu$, and does in general contain propagating modes. The second equation in \eqref{consAb} states that $A_\mu$ has constant norm with respect to the hatted metric and excludes all of the previously-mentioned propagating solutions. The system therefore describes a locally $AdS_3$ spacetime together with one of its constant norm self-dual vectors.  
Since the equations \eqref{consAb} are  non-linear,
 it is not entirely clear what is the structure, or even the number, of solutions. 

The second way to solve for $A_\mu$ 
makes use of \eqref{killv}. 
AdS$_3$ has six Killing vectors, and the self-duality condition selects three of them. $A_\mu$ is then a linear combination of these three Killing vectors, subject to the constant norm condition $\hat A^2=1-e^{-4U}$. This would seem to imply that we obtain a two-parameter family of solutions. Nevertheless, one can check explicitly (e.g. by working in Poincar\'e coordinates and using the  transformation formulae given in \cite{Maldacena:1998bw}) that all solutions in this family are related by $SL(2,\mathbb{R})_R$ transformations, which leave the AdS$_3$ metric $\hat g_{\mu\nu}$ invariant, but change the reference self-dual Killing vector.



\subsubsection*{The NHEMP truncation}

In the case of the NHEMP truncation, we can similarly concentrate on a slice through the phase space that has  $U_{1,2}$ constant. As before, this condition is sufficient to suppress all the propagating degrees of freedom. We find it convenient to let again the metric take the form \eqref{defgh}, with $\hat g_{\mu\nu}$ defined as in \eqref{eqhatg}, where the new scalar $U$ is fixed by the relative normalization between the curvature of $g_{\mu\nu}$ and $\hat g_{\mu\nu}$. The scalar equations of motion \eqref{eq3} - \eqref{eq4} then imply that the norms of $A_2$ and $A_1-A_2$ are constant. A natural Ansatz (and likely the only solution) is to let $A^{1,2}_{\mu} = \alpha_{1,2} A_\mu$   where $A_\mu$ obeys \eqref{consAb}. Plugging this into the equations of motion, we find that the values of the two scalars are not independent, but are related as 
\be
e^{2U_2}=2e^{2U_1}-e^{-2U_1}.
\ee
Then $\alpha_{1,2}$ are given by \eqref{solsa} and the scalar $U$ is given by
%
%
\be
e^{4U} = e^{6U_1+2 U_2} = 2 e^{8U_1} - e^{4U_1}.
\ee
Note for further use that the relations \eqref{solsa} imply 
\be
2 \alpha_1 \alpha_2 - \alpha_2^2 = 1.
\label{propa}
\ee 
Therefore, we have managed to reduce the problem of finding the most general non-propagating modes in the NHEMP truncation to the same equations \eqref{defgh}, \eqref{eqhatg} and \eqref{consAb} as we had before. 

\subsubsection*{Comments and clarifications}

One important solution of the system of equations \eqref{defgh}-\eqref{consAb} is the case when $\hat g_{\mu\nu}$ is the Poincar\'e $AdS_3$ metric and $\hat A^\mu$ is one of its self-dual Killing vectors. We then recognize \eqref{defgh} as the standard construction of spacelike, null and timelike  warped $AdS_3$; the type of warping depends on the norm of the self-dual Killing vector $\hat A^\mu$  \cite{Coussaert:1994tu}, which in turn depends on the sign of $U$. The  minus sign in \eqref{defgh} indicates that the $S^1$ fiber of the warped $AdS_3$ spacetime is squashed
\footnote{If one starts instead with global $AdS_3$, which has a compact angular coordinate, closed timelike curves  appear at large radius in the warped metric. Such spacetimes are also known as three-dimensional G\"odel spacetimes \cite{Godel:1949ga,Reboucas:1982hn,Banados:2005da}. Closed timelike curves in squashed $AdS_3$ can be avoided by decompactifying the angular direction. 
 All known string constructions of warped $AdS_3$ require such an uncompactified setting.},  in the language of \cite{Anninos:2008fx}.
 In our conventions, the warping breaks the symmetry group of the original $AdS_3$   as
\be
SL(2,\mathbb R)_L \times SL(2,\mathbb R)_R \rightarrow SL(2,\mathbb R)_L \times U(1)_R.
\ee
Applying diffeomorphisms to the resulting metric, one can span the entire solution space of the system, such as the black string sector.

\bigskip

\noindent Next, we would like to discuss a  possibly confusing point, taking the S-dual dipole theory as an example. We have shown that on each constant $U$ slice through the solution space of the S-dual dipole theory (here $U$ is completely arbitrary), one can build  solutions starting from an arbitrary AdS$_3$ metric $\hat g_{\mu\nu}$. This metric could, in particular, be that of the BTZ black string, characterized by the left/right-moving temperatures $T_\mp$,
\be
d\hat s^2_{BTZ} = \ell^2 \left( T_+^2 d v^2 + 2  r d v du +  T_-^2  du^2   +\frac{ d r^2}{ 4( r^2 - T_+^2 T_-^2)}  \right).
\label{btzmet}
\ee
This is of course the seed metric for the construction of the warped black strings \eqref{bsmet}. Nevertheless, from  \eqref{Abh} it seems that fixing the value of $U$  will fix $T_+$, thus restricting the possible auxiliary metrics $\hat g_{\mu\nu}$ one can choose.

This argument is however not correct, and it is easy to see why: the relationship between $U$ and $T_+$ involves specifying the parameter $\l$, which from the point of view of the classical solution space does not have any meaning. To understand this, it may be a useful exercise to work out explicitly the solution for $A_\mu$, when $\hat g_{\mu\nu}$ is the BTZ metric \eqref{btzmet}. The solution is given by a linear combination of the three self-dual Killing vectors of $\hat g_{\mu\nu}$
\be
A = \a_{-1} \s_{-1} + \a_0\, \s_0 + \a_1 \s_1
\ee
where\footnote{The Lie bracket algebra of these one-forms is
$[\s_{\pm 1},\s_0] = \mp \frac{2}{\ell} \s_{\pm 1}\,, \; [\s_1,\s_{-1}] = - \frac{2}{\ell} \s_0$.
}
\be
\s_{-1} = \frac{\ell}{2T_+}\sqrt{r^2 - T_+^2 T_-^2} \left(2  du + \frac{T_+  dr}{r^2 - T_+^2 T_-^2}\right) \, e^{-2 T_+ v} \;, \;\;\;\;\; 
\s_0 = \frac{\ell}{T_+}\,( r du + T_+^2 dv)
\ee
\be
\s_1 =\frac{\ell}{2T_+}\sqrt{\frac{r-T_+ T_-}{r+ T_+ T_-}} \left( (r+T_+ T_-) du  - \frac{T_+ dr}{2 (r-T_+ T_-)}\right) \, e^{2 T_+ v}
\ee
which obey $\s_{-1}^2 = \s_1^2 = 0, \, \, \s_0^2 = 1$. The norm of the vector $A_\mu$ is
\be
\hat A^2 =   \a_0^2 - 2 \a_{-1} \, \a_1.
\ee
It should now be clear that for each choice of $U$, one can find solutions to $\hat A^2 = 1-e^{-4U}$ for any $T_\pm$, by choosing the $\a_i$ appropriately. 
As explained earlier in this section, one can always choose a particular representative in this family of solutions; all the others are related to it by $SL(2,\mathbb{R})_R$ conformal transformations. A natural choice is to look for a
%
stationary ($v$ independent) representative.  It is clear that a stationary solution exists for any $U>0$ and any $T_\pm$, since we can choose $A \propto \s_0$, which is $v$ independent. Nevertheless, when $U=0$ and thus $\hat A^2=0$, one cannot find a time-independent solution, as long as $T_+ \neq 0$; therefore, stationary solutions with $U=0$  require $T_+=0$. 
  When $U<0$, stationary solutions can still be obtained, but only in the sector $T_+^2 <0$.

Note that the stationary solutions $A = \a_0 \, \s_0$ that we obtain for $U>0$ are not smooth as $T_+ \r 0$. To make them smooth, one should introduce a parameter $\l$, with dimensions of length, such that
\be
\lim_{T_+ \r 0} \, \frac{\a_0}{T_+} = \l \;,\; \;\;\;\;\; \a_0 = \sqrt{1-e^{-4U}}.
\ee
A particular functional form of $\a_0$ that  satisfies this requirement is 
\be
\a_0 = \frac{\l T_+}{\sqrt{1+\l^2 T_+^2}},
\ee
which corresponds to the value of $U$ in \eqref{Abh}. Although its introduction in the classical solution is just a matter of convenience, $\l$ does have a physical meaning from the point of view of holography: it represents the intrinsic length scale  of the holographic dual to warped AdS$_3$, which is a dipole theory \cite{Bergman:2000cw}.  In the holographic interpretation of the warped black strings given in \cite{decrypt}, it is explained why all fields in the gravitational description, including the scalar $U$, are naturally expressed in terms of $\l T_+$. 




\subsection{Equivalence of the warped and unwarped symplectic structures}
\label{equiv}

In the previous section, we showed that there exists a universal subsector of both the S-dual dipole and NHEMP truncations that is entirely characterized by a negative curvature Einstein metric, together with one of its constant norm self-dual vectors. Given the intimate relation between the space of solutions of a theory and its phase space, we may expect that this  subsector of the phase space of the truncations is in fact
 \emph{isomorphic} to the phase space of pure three-dimensional Einstein gravity with a negative cosmological constant.

In this section, we will  prove that this is indeed the case, by showing that the symplectic form of the S-dual dipole and NHEMP truncations on the constrained solution space is \emph{identical} to the symplectic form of pure Einstein gravity in AdS$_3$.  
The mapping between symplectic forms will require  fixing the otherwise ambiguous boundary term \eqref{ambiom} in the symplectic structure in a specific manner.

\subsubsection*{S-dual dipole theory}


We will concentrate our attention on the presymplectic form $\boldsymbol \Theta$ of the S-dual dipole theory, restricted to the constant $U$  slice through the phase space, where $g_{\mu\nu}$ and $A_\mu$ obey \eqref{defgh}, \eqref{consAb} and \eqref{killv}. The  presymplectic form receives contributions from the Einstein and Chern-Simons terms in \eqref{psf}; the scalar contribution vanishes because we are on a slice of constant $U$. 

As explained in section \ref{symp}, equation \eqref{ambig}, the presymplectic form is ambiguous up to the addition of an exact form $\boldsymbol d \boldsymbol Y$. The statement that we will prove below is that there exists a choice of boundary term $\boldsymbol Y$, such that the total presymplectic form, when expressed in terms of $\hat g_{\mu\nu}$ and $A_\mu$ using  \eqref{defgh}-\eqref{consAb}-\eqref{killv}, only depends on  $\hat g_{\mu\nu}$ and, moreover, it precisely equals the presymplectic form of pure three-dimensional Einstein gravity with metric $\hat g_{\mu\nu}$. Formally, we show that
\bea
\boldsymbol \Theta_g[ g,\delta  g]  + \boldsymbol \Theta_{CS}[A,\delta A] +\boldsymbol d \boldsymbol Y[ g ,A,  \d g, \delta A] = \boldsymbol \Theta_{\hat g}[\hat g,\delta \hat g]   \label{sum}
\eea
where\footnote{By an abuse of notation, we have denoted by $\boldsymbol Y$ both the ambiguity \eqref{ambig} in the symplectic form and the ambiguity above. The two differ by a factor of $16 \pi G_3$, due to the definition \eqref{thsplit}.
}
\be
\boldsymbol Y 
= - \eps_{\mu \alpha \beta} A^\alpha \delta  A^\beta  dx^\mu ,\label{defTB}
\ee
and the variations of the fields must obey the linearization of the constraints \eqref{defgh} - \eqref{consAb}. Here, the variation $\d A^\mu$ is defined as
\bea
 \d A^\mu \equiv  \d ( A^\mu )=  g^{\mu\nu} \d A_\nu + \delta g^{\mu\nu}  A_\nu .
\eea
We now proceed to proving this statement.

Let  $\hat h_{\mu\nu} = \delta \hat g_{\mu\nu}$ be an arbitrary variation of the metric $ \hat g_{\mu\nu}$, which on the constant $U$ slice leads to a variation $h_{\mu\nu} = \delta g_{\mu\nu}$ of $g_{\mu\nu}$. The two are related by
\be
h_{\mu\nu} = e^{4U} (\hat h_{\mu\nu} - A_\mu \d A_\nu - A_\nu \d A_\mu).
\ee
Indices on  $h_{\mu\nu}$, $\hat h_{\mu\nu}$ are raised  with $g^{\mu\nu}$ and respectively $\hat g^{\mu\nu}$. The latter are related via
\be
g^{\mu\nu} = e^{-4U} \hat g^{\mu\nu} + \hat A^\mu \hat A^\nu \;, \;\;\;\;\; \hat A^\mu = A^\mu.
\ee
%
%
The constraint \eqref{consAb} implies that the variation of $A_\mu$ must satisfy
\be
2 \hat A^\mu \d A_\mu - \hat A^\mu \hat h_{\mu\nu} \hat A^\nu =0.
\ee
Other useful relations involving the perturbations are
\be
 \d \hat A^\mu \equiv  \d ( \hat A^\mu )=  \hat g^{\mu\nu} \d \hat A_\nu - \hat h^{\mu\nu}  \hat A_\nu\;,\qquad \hat h = h,
\ee 
 \be
 h^{\mu\nu} = e^{-4U} \hat h^{\mu\nu} - \d \hat A^\mu \hat A^\nu - \d \hat A^\nu \hat A^\mu  .\label{delAup}
\ee
Finally, the Christoffel symbols for $g_{\mu\nu}$ are given in terms of those for $\hat g_{\mu\nu}$ via
\be
\Gamma^\l_{\mu\nu} = \hat \Gamma^\l_{\mu\nu} + \half (\hat F^\l{}_\mu A_\nu + \hat F^\l{}_\nu A_\mu)- \half \, e^{4U} \hat A^\l (\hat \nabla_\mu A_\nu + \hat \nabla_\nu A_\mu) \;, \;\;\;\;\; \Gamma^\l_{\l \mu} = \hat \Gamma^\l_{\l\mu}.
\ee
Note that the last term in $\Gamma^\l_{\mu\nu}$ can be dropped, since $A_\mu$ is a Killing vector of $\hat g_{\mu\nu}$. 

To compute $\boldsymbol \Theta_g$ in \eqref{psf}, we need the expressions for  
\bea
\nabla_\s h^{\mu\s}&=& e^{-4U} \hat \nabla_\s \hat h^{\mu\s} + \hat A^\s \hat A^\l \left(\nabla_\s \hat h^\mu{}_\l - \half \hat \nabla^\mu\hat h_{\s\l} \right) + \half \hat A^\mu \hat A^\l \p_\l \hat h +\nonumber \\
&& \hspace{1 cm } + e^{-4U} \hat h^{\l \s} \hat F^\mu{}_\s A_\l   - \hat F^\mu{}_\s \d \hat A^\s (1-e^{-4U}), \nonumber \\ \nonumber\\ 
\nabla^\mu h &=& e^{-4U} \hat \nabla^\mu \hat h + \hat A^\mu \hat A^\nu \hat \nabla_\nu \hat h.
\eea
The contribution to $\boldsymbol \Theta_{CS}$ is
\be
\Theta^\mu_{CS} =  -\frac{2}{ \ell} \, \e^{\mu\nu\rho} A_\rho \d A_\nu = -\, e^{-4U} \hat F^{\mu\nu} \d A_\nu.
\ee
In terms of $\hat g_{\mu\nu}$ and $A_\mu$, the boundary term $Y_\mu$ reads
\be
Y_\mu = -  \e_{\mu\a\b} A^\a \d A^\b = - e^{4U} \hat \e_{\mu\a\b} \hat A^\a \d \hat A^\b
\ee
where we have used the fact that $\sqrt{-g} = e^{4U} \sqrt{-\hat g}$. Next, using  \eqref{Hodge} we find
\be
\left( \boldsymbol d \boldsymbol Y \right)^\mu= -\hat \nabla_\lambda (\hat A^\mu \delta \hat A^\lambda - \hat A^\lambda \delta \hat A^\mu).
\ee
Using the fact that $\hat \nabla_\mu \hat A^\mu = \hat A^\l \hat \nabla_\l A_\mu =0$ (as well as the  variations of these constraints), we can show that
\be
\hat \nabla_\l (\hat A^\mu \d \hat A^\l - \hat A^\l \d \hat A^\mu ) = 2 \,\d \hat A^\l \hat \nabla_\l \hat A^\mu -\half \hat A^\mu \hat A^\s \ \p_\s \hat h+  \,\hat A^\l \hat A^\s ( \hat \nabla_\l \hat h^\mu{}_\s  -\half \hat \nabla^\mu \hat h_{\l\s}).
\ee
Summing up all the contributions, we find the identity \eqref{sum}.

\subsubsection*{NHEMP truncation}

The proof of \eqref{sum} for the case of the NHEMP truncation proceeds in an identical manner. Since we have parametrized the relationship between $g_{\mu\nu}$ and $\hat g_{\mu\nu}, A_\mu$ in the same way as for the dipole theory (in particular, the field $A_\mu$ satisfies the same equations in terms of the auxiliary metric $\hat g_{\mu\nu}$ and $U$), all that we need to check is that the Chern-Simons contribution to the presymplectic form given in \eqref{psf}, when written in terms of $A_\mu$, takes the same form as for the dipole theory. Using  \eqref{nhekdata},  we find
\be
\Theta^\mu_{CS} = c_{ij} \e^{\mu\nu\rho} A_\rho^i \d A_\nu^j 
 =  
-\frac{2}{\ell} \, \e^{\mu\nu\rho} A_\rho \, \d A_\nu (2 \a_1 \a_2 - \a_2^2) =- \frac{2}{\ell} \, \e^{\mu\nu\rho} A_\rho \, \d A_\nu
\ee
where we plugged in  $A_{1,2} = \alpha_{1,2} A$ and then used the property \eqref{propa}. Therefore, the Chern-Simons contribution is exactly identical to the one in the S-dual dipole theory. Consequently, we can use exactly the same boundary contribution $\boldsymbol Y$ as \eqref{defTB} in order to prove  the equivalence \eqref{sum}. A natural way of rewriting this contribution in terms of the fields $A_{1,2}$ appearing in the NHEMP truncation action is 
\be
 Y_\mu 
=\frac{\ell}{2}\, c_{ij} \, \e_{\mu\alpha \beta}\, A_i^\a \, \d A_j^\b  \label{Ygen}
\ee
where we have again used \eqref{propa}. The presymplectic form of the NHEMP truncation on the constraint surface $U_{1,2} =$ constant is therefore identical to the one of pure Einstein gravity. 

\bigskip

The fact that the total presymplectic form on the constraint surface is the same as the presymplectic form of Einstein gravity implies that the symplectic form must also be the 
same, since it is simply given by \eqref{sympl}.
This mapping of symplectic structures means that all dynamical quantities in one theory can be mapped to analogous quantities in the other theory. This provides us with a powerful equivalence, which can be used to construct of a phase space for asymptotically warped $AdS_3$ spacetimes for each known choice \cite{Brown:1986nw,Compere:2013bya,Troessaert:2013fma} (see also \cite{Avery:2013dja,Apolo:2014tua}) of boundary conditions for Einstein gravity in $AdS_3$.

 Furthermore, since the symplectic form is related to the conserved charges via \eqref{chom1}, we find that there is a \emph{one-to-one correspondence} between conserved charges in warped AdS$_3$ and in pure Einstein gravity in AdS$_3$. More precisely, for any consistent choice of boundary conditions in AdS$_3$ that leads to charges that are finite, integrable and conserved, we obtain a corresponding set of boundary condition in warped AdS$_3$ with \emph{identical} charges. In particular, the charges are completely independent of the value of the scalar $U$ - a feature of our construction which has been built into the definition of $\hat g_{\mu\nu}$ in \eqref{defgh}.  In the next subsections, we 
will explicitly construct two such phase spaces and their symmetries, obtained from Dirichlet (also known as Brown-Henneaux) \cite{Brown:1986nw} and Dirichlet-Neumann chiral boundary conditions \cite{Compere:2013bya} in the auxiliary $AdS_3$.

\subsection{Dirichlet boundary conditions}
 \label{excon}


The most general non-linear solution to the three-dimensional Einstein equations with negative cosmological constant $ -1/\ell^2$ can be written in Fefferman-Graham form as
\be
d\hat s^2 =  \ell^2  \rho^2\, \left(  \hat g^{(0)}_{ij} + \rho^{-2} \hat g^{(2)}_{ij} + \rho^{-4} \hat g^{(4)}_{ij}\right) dx^i dx^j + \ell^2 \frac{ d\rho^2}{\rho^2}.\label{FG}
\ee
It is a peculiarity of three dimensions that the expansion terminates at finite order \cite{Skenderis:1999nb}. The  various coefficients are related as
\be
\hat \nabla^a \hat g^{(2)}_{ab} = \hat \nabla_b \hat g^{(2) c}_c\;, \;\;\;\;\; \hat g^{(2) c}_c =-\frac{1}{2}  R[\hat g^{(0)}]\;, \;\;\;\;\; \hat g_{ab}^{(4)} =\frac{1}{4}\, \hat g_{ac}^{(2)} \,\hat g^{(0)\, cd}\, \hat g_{db}^{(2)}.
\ee
Given such a metric, the solution for $A_\mu$ satisfying \eqref{consAb}  admits an infinite expansion in $1/\rho$, the first few terms of which were given in \cite{fg}. 

Dirichlet or Brown-Henneaux boundary conditions \cite{Brown:1986nw} consist in imposing
\be
\hat g_{ab}^{(0)}dx^a dx^b = -dt^+ dt^- \label{fbm}
\ee
where  we take $t^\pm$ to be non-compact. Then, the solution for $\hat g^{(2)}_{ab}$ is
\be
\hat g_{ab}^{(2)}dx^a dx^b = \frac{1}{k} L(t^-) (dt^-)^2+\frac{1}{k} \bar L(t^+) (dt^+)^2\;,\; \;\;\;\;\; k \equiv \frac{\ell}{4G_3} \label{dform}
\ee
where $L(t^-)$, $\bar L(t^+)$ are two arbitrary functions of their respective arguments which are allowed to vary on the phase space. General boundary conditions can then be obtained by acting on the solution space with trivial diffeomorphisms, i.e. diffeomorphisms whose linearized generators are associated with  conserved charges that are identically zero. Since trivial diffeomorphisms do not contain any physics, we restrict our discussion to the Fefferman-Graham gauge. 

The  diffeomorphisms that leave the asymptotic metric unchanged take the form
 \bea
\xi^\mu \p_\mu &=&\left( \eps_+(t^+) +\frac{1}{2\rho^2}\eps_-''(t^-) +\O(\rho^{-4})\right) \, \p_+ +\left( \eps_-(t^-) +\frac{1}{2\rho^2}\eps_+''(t^+) +\O(\rho^{-4})\right)\, \p_- + \nonumber\\
&&+ \rho \left( -\frac{1}{2}\eps_+'(t^+) -\frac{1}{2}\eps_-'(t^-)  +\O(\rho^{-1})\right)\, \p_\rho. \label{AKV}
\eea
%
These asymptotic Killing vectors satisfy a Lie bracket algebra which consists of two copies of the Virasoro algebra. 
The associated conserved charges represent the Lie bracket algebra by a Dirac algebra, which now consists of two copies of the centrally-extended Virasoro algebra, with central charges $c_L=c_R = 6k$. 

We would now like to find the solution for the vector field $A_\mu$, which satisfies \eqref{consAb} and \eqref{killv}. Since $\hat A^\mu$ is a Killing vector of the spacetime \eqref{FG} - \eqref{dform}, asymptotically it must be of the form \eqref{AKV}.
At leading order in $1/\rho$, the self-duality condition requires that $\eps_-(t^-) = 0$, while the constant norm condition imposes
\be
-\frac{1}{2}\eps_+'' \eps_+ + \frac{1}{4}(\eps_+')^2 + \frac{\bar L}{k}\eps_+^2 = 1-e^{-4U}  \equiv \chi  . \label{EQ1}
\ee
A convenient way to solve this constraint consists in setting 
\be
\eps_+ =  e^{-\Phi(t^+)}\label{defe}
\ee
which implies 
\be
\bar L = \frac{k}{4} \left( (\p_+\Phi)^2 -2 \p_+^2 \Phi +4 \chi  e^{2\Phi} \right).\label{barL}
\ee
This expression is nothing else than the stress-tensor of a chiral right-moving boson with a Liouville potential.  This is a familiar concept from the Hamiltonian reduction of the Chern-Simons formulation of Einstein gravity \cite{Coussaert:1995zp}. Here, the same structure appears in the metric formalism as well. Note that in terms of $\bar L$, the solution for $\Phi$ is entirely non-local. 

Moving on to asymptotic symmetries, $L$ transforms under left-moving diffeomorphisms $\e_- = f_-$ according to the usual law
\be
\delta_{f_-} L  = f_- \p_- L +2 L \p_- f_- - \frac{k}{2}\,\p_-^3 f_- . \label{eq11}
\ee
Under a right-moving diffeomorphism \eqref{AKV} of generator $\eps_+ = f_+$,  the commutator between asymptotic Killing vectors implies the following transformation law for $\Phi$ 
\be
\delta_{f_+} \Phi = f_+ \p_+ \Phi + \p_+ f_+ ,\label{eq9}
\ee
which in turn implies, using the definition \eqref{barL}, the usual Virasoro transformation law
\bea
\delta_{f_+} \bar L &=& f_+ \p_+ \bar L +2 \bar L \p_+ f_+ - \frac{k}{2}\p_+^3 f_+.\label{eq10}
\eea
Equations \eqref{AKV}-\eqref{defe} give the leading terms in the asymptotic expansion of the Killing vector $\hat A^\mu$ in terms of the holographic data $\bar L$ \eqref{barL}. To find the full solution, we find it convenient  to work in  the gauge
\be
g_{\rho\rho} = \frac{\ell^2}{\rho^2}\;,\qquad g_{\rho -} =0\;,\qquad A_\rho = 0,\label{gc}
\ee
in which it takes a rather simple form, as we will soon see. To reach this gauge, one can perform a trivial coordinate transformation 
\be
t^+ \rightarrow  t^+  +\O(\rho^{-4}),\qquad  t^- \rightarrow  t^- - \frac{\p_+ \Phi }{2  \rho^2}+\O(\rho^{-6}),\qquad  \rho \rightarrow  \rho.
\ee
In this gauge, the  $AdS_3$ metric  and the massive vector field are given  by 
\bea
\frac{d\hat{s}^2}{\ell^2} &=&   \frac{d\rho^2}{\rho^2} - \frac{\p_+ \Phi}{\rho} \, d\rho dt^+ +\left( \frac{1}{4}(\p_+ \Phi)^2+ \chi\, e^{2\Phi} \right) (dt^+)^2 
  -\left(\rho^2 +\frac{\chi e^{2\Phi} L}{k  \rho^2}\right) dt^+ dt^- +\frac{L}{k}(dt^-)^2,\nn \\
\frac{A}{  \ell} &=&  \chi \, e^\Phi dt^+ - \left(\frac{ e^{-\Phi}}{2}\rho^2 + \frac{ \chi e^\Phi L}{2 k  \rho^2}\right) \, dt^- \label{gfA}
\eea
where $\chi$ is related to the constant scalar $U$ via \eqref{EQ1}. Quite remarkably, the radial expansion of $A_\mu$ exactly stops, so we have the full non-linear solution to the system \eqref{consAb} - \eqref{killv}, including all boundary gravitons. The warped AdS$_3$ metric constructed using \eqref{defgh} then also admits a finite radial expansion, given by 
\bea
(1-\chi ) \frac{ds^2}{ \ell^2} &=&   \frac{d\rho^2}{\rho^2} - \frac{\p_+\Phi}{\rho}d\rho dt^+ + \left( \frac{1}{4}(\p_+ \Phi)^2+\chi(1-\chi)\,e^{2\Phi} \right)(dt^+)^2  \nonumber \\
&&\hspace{-2.5cm} -(1-\chi) \left(  \rho^2 +\frac{ \chi e^{2\Phi} L}{k  \rho^2}\right) dt^+dt^- - \left( \frac{1}{4} e^{-2\Phi}\rho^4 - (1-\frac{\chi}{2} )\frac{L}{k}+\frac{\chi^2 e^{2\Phi} L^2}{4k^2 \rho^4} \right)(dt^-)^2 .\label{ps}
\eea
This is the main result of this section. Given that, in this particular gauge, the expansion exactly stops for Dirichlet boundary conditions, it  is  interesting
to ask  whether the gauge conditions \eqref{gc} can be useful for constructing the full solution with a general boundary metric, or 
when propagating modes are present.

It is not hard to see that the black string solutions  agree with the above metric for constant $\Phi, L$ and $\bar L$ if we identify
\be 
L = k T_-^2\, ,\;\;\;\qquad e^\Phi = \frac{1}{\sqrt{\chi}} T_+ \;\; \Rightarrow \;\; \bar L = k T_+^2
\ee
using \eqref{barL}. The solution then reduces to the warped black strings \eqref{bsmet}-\eqref{Abh} upon identifying
\be
 v = t^+\;, \;\;\;\;\; u=t^-\;, \;\;\;\;\; \rho^2 = -2r+2\sqrt{r^2 - T_-^2 T_+^2}
\ee 
and parametrizing $U$ as in \eqref{Abh}. 

By construction, the phase space admits as asymptotic symmetry algebra two copies of the Virasoro algebra. More precisely, in the coordinates that we are using, the phase space is preserved by the action of the following asymptotic symmetry generators
\bea
\xi &=& \left( f_+ +\frac{\p_-^2 f_-}{2\rho^2}+O(\rho^{-6}) \right)\p_+ + \left( f_- +\frac{\chi e^{2\Phi}\p_-^2 f_-}{2\rho^4} +O(\rho^{-8}) \right)\p_-  \nonumber\\
&&+ \left( -\frac{\rho}{2}(\p_- f_- + \p_+ f_+) +\frac{\p_+\Phi \p_-^2 f_-}{4\rho}+O(\rho^{-5})\right) \p_\rho\, \label{aKV}
\eea
where $f_+ = f_+(t^+), f_-=f_-(t^-)$. The transformation laws of the fields $\Phi(t^+)$, $\bar L(t^+)$ and $L(t^-)$ are given by \eqref{eq11}-\eqref{eq9}-\eqref{eq10}.

Given the phase space \eqref{ps} and the asymptotic Killing symmetries \eqref{aKV}, one can then compute the conserved charges using the explicit expressions \eqref{ecc} and the prescription for $\boldsymbol Y$ given in \eqref{defTB}\footnote{A subtlety is that the asymptotic Killing vectors \eqref{aKV} are field-dependent, $\xi=\xi[g]$, while the variation in \eqref{consch} is defined with $\delta \xi =0$. One can alternatively define $\delta$ as acting on all the fields, case in which the charge is given by $\boldsymbol k_\xi[\delta \phi]= \d \boldsymbol  Q_\xi [\phi]  -\boldsymbol  Q_{\delta \xi} [\phi]  - \xi \cdot \boldsymbol  \Theta [\phi,\d\phi] $.}. Since we proved the equivalence of symplectic forms \eqref{sum} on the constraint surface $ U= $ constant, we are guaranteed that the resulting charge one-forms will be given by the usual expressions in AdS$_3$, 
\bea
j^\xi &=&\int_{\bar{\hat g}}^{\hat g} k^\xi[\delta \hat g] =  \frac{1}{2\pi}  \left( - f^+ \bar L dt^+ +  f^- L dt^- \right) ,\label{chc}
\eea
where $\bar{\hat g}_{\mu\nu}$ is the Poincar\'e AdS$_3$ background. We can also  use the explicit expression for the charges to check that the above expression is true.
 These charge currents then represent two commuting copies of the Virasoro algebra, with the usual central charges $c_L = c_R = 6k$.

Note that the scalar $U$ is held fixed when varying $\hat  g_{\mu\nu} $ in  \eqref{chc}. This makes sense from a holographic point of view since, as shown in \cite{decrypt}, $U$ is related to the source for an irrelevant operator in the dual theory and should thus be kept fixed. Nevertheless, by construction, the charges \eqref{chc} do not depend upon $U$. 
Therefore, from the point of view of the covariant phase space construction, $U$ should be allowed to vary, as long as it is a constant. A way to incorporate a varying 
 constant $U$ in the 
 holographic picture is to note that
  the irrelevant operator that has $U$ as coefficient can also be interpreted as an exactly marginal operator from the non-relativistic point of view \cite{Guica:2010sw,decrypt}. In this case,  $U$ becomes a modulus, which may then be allowed to vary on the phase space.

The construction of this section was performed at full non-linear level in the gauge \eqref{gc}. We find it useful nevertheless to repeat this construction using radial gauge for the auxiliary AdS$_3$ spacetime, in which the holographic interpretation of the various metric components is the clearest. Thus, in appendix \ref{linsol}, we 
present the explicit construction of the 
 linearized phase space for non-bulk-propagating perturbations around null warped AdS$_3$ in radial gauge.

\subsection{Dirichlet-Neumann chiral boundary conditions\label{dnbc}}

Let us now construct the phase space for asymptotically warped $AdS_3$ spacetimes that can be obtained from  mixed Dirichlet-Neumann chiral boundary conditions in the auxiliary $AdS_3$ \cite{Compere:2013bya}. We will obtain a phase space similar to the ones first constructed in \cite{Compere:2009zj,Anninos:2010pm} for the case of topologically massive gravity (see also \cite{Blagojevic:2009ek,Henneaux:2011hv,sss}). 

Note that this problem involves  two chirality choices:  one in the boundary conditions \cite{Compere:2013bya} and the second, in the  choice of orientation of the self-duality condition \eqref{kv}. One can therefore align or anti-align these chiralities, and the result will differ. We will discuss here only one choice of interest; the other
can be analysed in a similar fashion and  will  be briefly commented upon at the end of this section. 

The general solution to Einstein's equations with a negative cosmological constant is given by \eqref{FG}. Chiral boundary conditions at fixed real $\bar \Delta$ consist in imposing \cite{Compere:2013bya} 
\bea
\hat g_{ab}^{(0)}dx^a dx^b &=& -dt^- (dt^+ - \p_-  P(t^-) dt^-),\non \\
\hat g_{ab}^{(2)}dx^a dx^b &=& \frac{\bar \Delta}{k} (dt^+ - \p_- P(t^-)dt^- )^2+\frac{1}{k} L(t^-) (dt^-)^2 \label{hatg2}
\eea
where  $k$ is  given by \eqref{dform} and
 $P(t^-)$, $L(t^-)$ are two arbitrary functions, which are allowed to fluctuate. They represent the boundary photons and, respectively,  gravitons. 
 These boundary conditions are characterized by the constant value $g_{++}=\frac{\bar \Delta}{k}\ell^2$ or, equivalently, by the constant right-moving zero mode
\bea
\mathcal Q_{\p_+} = \bar\Delta.\label{fixedDelta}
\eea
 The diffeomorphisms that leave the asymptotic metric unchanged take the form
\bea
\xi^\mu \p_\mu &=&\left( \sigma(t^-) 
+\O(\rho^{-2})\right) \, \p_+ + \eps_-(t^-)  \p_- + \rho \left(  -\frac{1}{2}\eps_-'(t^-)  +\O(\rho^{-1})\right)\, \p_\rho. \label{ask1}
\eea
We are interested in solving the equations \eqref{consAb} for $A_\mu$ when the hatted metric takes the form  \eqref{hatg2}. Note that this metric can be obtained from \eqref{fbm} - \eqref{dform} via the coordinate transformation $t^+ \r t^+ - P(t^-)$ and setting $\bar L (t^+) = \bar \Delta$. Therefore, the solution for $A_\mu$ is the same as before \eqref{gfA}, up to the above gauge transformation and fixing $\bar L (t^+)$. Just as before, we can find a solution for $A_\mu$ for every set of boundary data $L(t^-), \, \bar \Delta$ and any value of the scalar $U$. Note that $A_\mu$ need not take the asymptotic form  \eqref{ask1}, since only four of the six Killing vectors of AdS$_3$ are of this form.


The final solution for the warped AdS$_3$ metric, vector and scalar fields is
\bea
 \frac{ds^2}{ \ell^2} &=&   \frac{d\rho^2}{(1-\chi )\rho^2} - \frac{\p_+\Phi}{(1-\chi )\rho}d\rho (dt^+-\p_- P dt^-) + \left( \frac{(\p_+ \Phi)^2}{4(1-\chi )}+\chi \,e^{2\Phi} \right)(dt^+-\p_- P dt^-)^2  \nonumber \\
&& - \left(  \rho^2 +\frac{ \chi e^{2\Phi} L}{k  \rho^2}\right)  (dt^+-\p_- P dt^-)dt^- - \left( \frac{1}{4} e^{-2\Phi}\rho^4 - (1-\frac{\chi}{2} )\frac{L}{k}+\frac{\chi^2 e^{2\Phi} L^2}{4k^2 \rho^4} \right)\frac{(dt^-)^2}{(1-\chi )} ,\nn\\
\frac{A}{  \ell} &=&  \chi \, e^\Phi (dt^+-\p_- P dt^-) - \left(\frac{ e^{-\Phi}}{2}\rho^2 + \frac{ \chi e^\Phi L}{2 k  \rho^2}\right) \, dt^- ,\qquad e^{2U} = \frac{1}{\sqrt{1-\chi}}. \label{chiralsol}
\eea
The function $\Phi(t^+)$ is given by \eqref{barL} in terms of $\bar\Delta$   and $U$. We can obtain solutions with 
 $\Phi(t^+) = \Phi=$ constant in three different cases: $U> 0 $ and $\bar \Delta > 0$; $U = 0$ and $\bar \Delta = 0$; $U < 0 $ and $\bar \Delta < 0$. Since $g_{++}= \frac{\bar\Delta \ell^2}{k}$, these solutions correspond to asymptotically spacelike, null or timelike warped $AdS_3$ spacetimes. 
 The above boundary conditions are the analogue for the S-dual dipole theory of the boundary conditions first found in \cite{Compere:2009zj,Anninos:2010pm} in the context of topologically massive gravity.  There, it was observed that the condition that  \eqref{fixedDelta} be fixed on the phase space is a natural requirement in order to define integrable charges. It was understood in \cite{Compere:2013bya}  that this is, more fundamentally, a sufficient condition in order for the action to admit a variational principle. Note also that since $\Phi$ is fixed, the leading term in the metric does not fluctuate under these boundary conditions.

The asymptotic symmetry algebra is identical to the one of \cite{Compere:2013bya}, by construction. This consists of a left-moving Virasoro algebra and a ``crossover" left-moving $U(1)$ current algebra, whose zero mode is the right-moving generator $\p_+$. The vector fields that generate these symmetries are given by \eqref{ask1}. 
Setting $\epsilon_- =e^{int^-}$, $\sigma = e^{i n t^-}$, we define $ \cL_n = Q_{\epsilon_-}$, $ \cJ_n = Q_{\sigma}$. The asymptotic symmetries generated by these charges
 enhance the  isometries of the null warped background as 
\be
SL(2,\mathbb R)_L \times U(1)_R \rightarrow Vir_L \times \widehat{U(1)}_R.
\ee
 The central charge and level of the current algebra are given by
\bea
c_L = \frac{3\ell }{2G},\qquad \bar k_{KM} = -4 \bar \Delta \, .\label{ccL}
\eea
The Kac-Moody level $\bar k_{KM}$ is positive around the timelike warped AdS$_3$ vacuum $\bar \Delta = -k/4$, but it becomes negative in the presence of a $\bar \Delta>0$ black string. This is indicative of the fact that the symplectic form of the phase space with mixed chiral boundary conditions is not positive definite, as it can be easily checked.

Finally, let us note that if we instead anti-align the choice of chiralities in  \cite{Compere:2013bya} and \eqref{kv}, we obtain the following asymptotic symmetry enhancement 
\bea
SL(2,\mathbb R)_L \times U(1)_R \rightarrow \widehat{U(1)}_L \times Vir_R
\eea
with central charges
\bea
c_R = \frac{3\ell }{2G},\qquad k_{KM} = -4  \Delta \, ,\label{cc2}
\eea
where $\mathcal Q_{\p_-} = \Delta$ is the fixed left-moving zero mode. Boundary conditions in warped $AdS_3$ spacetimes can therefore be defined, which admit this unusual Ka\v{c}-Moody-Virasoro algebra as asymptotic symmetry algebra.


\section{Including the bulk propagating modes}
\label{propmodes}

So far, we have succeeded in showing that warped AdS$_3$ can admit the simultaneous action of two Virasoro symmetries, which eluded previous attempts to find them. Nevertheless, the phase space we defined in section \ref{excon}, upon which these symmetries act, does not contain any bulk propagating modes, and thus does not evade the criticisms of \cite{Amsel:2009ev}. 
In this section, we will also include linear  bulk fields propagating around the warped black string backgrounds
and ask whether a consistent  phase space can be defined that contains both the Virasoro boundary gravitons and the propagating modes, the latter being subject to appropriate normalizable boundary conditions.

 Our conclusion strongly depends on the asymptotic behaviour of the bulk propagating modes. More precisely, if no travelling waves are present -- as is the case for the S-dual dipole truncation -- then we show that a phase space can always be defined, which contains the propagating modes and is acted upon by the two Virasoro symmetries. If this conclusion extends to non-linear level - as it most probably does - it amounts to a proof that, at least from a semiclassical, (super)gravity point of view, the dual theory to S-dual dipole warped AdS$_3$ is a two-dimensional CFT. When travelling waves are present -- as we will show is the case for the NHEMP truncation -- the subtleties involving these modes will not allow us to define a phase space satisfying the usual requirements of finiteness and conservation of the symplectic form, and we leave their analysis to future work.

This section has two parts. We start with a rather general discussion of the behaviour of linearized bulk modes around warped AdS$_3$ and point out the characteristic features of the spectra of the two truncations. Then, we compute the symplectic form on the phase space of the S-dual dipole theory  and show that it can be  consistently defined at  linear level.

\subsection{Behaviour of the linearized bulk solutions }

We consider linear fields propagating around the warped black string backgrounds \eqref{bsmet} with temperatures $T_\pm$ (which include the Poincar\'e/global null warped background as  special cases). We will be working in Fourier space, where the linear perturbations are given by
\be
\delta \phi^i (u,v,r) = e^{- i \om u - i \k v} \, \delta \phi^i(r).
\ee
The equations of motion for both the S-dual dipole and the NHEMP truncation can be reduced to a set of second order differential equations for the linear scalar field $\delta U$ (respectively $\delta U_1$) which depend upon a real parameter $\mu^2$
\be
\p_r [ (r^2-T_+^2 T_-^2) \, \p_r \delta U(r)] + \left( \frac{1}{4} -\mu^2 +\frac{ - 2 r \kappa \omega + T_+^2 \omega^2+T_-^2 \kappa^2  }{4(r^2 -T_+^2 T_-^2)}\right)\delta U(r) =0.\label{Master2}
\ee
The equation above takes exactly the same form as the equation of motion for a free massive scalar propagating around the BTZ black string \eqref{btzmet}  with the same temperatures. In that simple example, the parameter $\mu$ is related to the mass $m$ of the scalar via
\be
\mu = \half \sqrt{1+m^2 \ell^2}
\ee
and, for Dirichet boundary conditions on the field, to the  conformal dimension $\Delta$  of the dual operator\footnote{Note that the standard  range of masses for which both Dirichlet and Neumann boundary conditions are allowed for the usual Klein-Gordon norm is $-1 \leq m^2 \ell^2 < 0$, which corresponds to $0 \leq \mu < \frac{1}{2}$.}  by $\Delta = 1 + 2 \mu$ \cite{Witten:1998qj}. 

In general, the two solutions to \eqref{Master2} are hypergeometric functions. When $T_- = 0$, $T_+ \neq 0$, they are Whittaker functions, and  when $T_+ = T_- = 0$, they are Bessel functions. The  equation always has a regular singular point at infinity, and thus the behaviour of the solution as $r \r \infty$ takes the form
\be
\delta U(r) \sim A \, r^{- \half + \mu} + B \, r^{-\half - \mu} \label{asyu} .
\ee
 For $\mu \in \mathbb{R}$, we find the usual power-law fall-offs. For $\mu \in i \mathbb{R}$, we obtain the so-called travelling wave solutions, which can carry flux across the boundary of the spacetime and are related to near-horizon superradiance \cite{Bardeen:1999px}. 
The $\,\sim \,$ sign indicates the leading radial falloffs of the field, which are generally followed by terms suppressed in the $r \rightarrow \infty$ limit.

Equation \eqref{asyu} only gives the asymptotic behaviour of the scalar fields. The metric and gauge field perturbations have slightly different falloffs, which  can be reconstructed from the scalar field solution, as explained in appendix \ref{app:EOM} for the special case of the NHEMP truncation. The asymptotic behavior at large radius is the same for both the S-dual dipole and NHEMP truncations, and reads 

\be
h_{uu} \sim r^{\frac{3}{2} \pm \mu} \;, \;\;\;\;\;\;
h_{uv} \sim r^{\half \pm \mu} \;,\; \;\;\;\;\;h_{vv} \sim r^{-\half \pm \mu},  \nn
\ee

\be
\d A_u \sim r^{\half \pm \mu} \;,\; \;\;\;\;\; \d A_v \sim  r^{-\half \pm \mu} \;,\; \;\;\;\;\;\d A_r \sim r^{-\frac{3}{2} \pm \mu} \label{fall}
\ee
where we have chosen the radial gauge for the metric perturbation.

%
%

It is quite remarkable that a single hypergeometric equation, \eqref{Master2}, controls the dynamics of scalar perturbations of both warped $AdS_3$ \eqref{bsmet} and  $AdS_3$ spacetimes \eqref{btzmet}. All the dependence on the details of the theory, the specific couplings in the Lagrangian and on the warping parameter $\lambda$ is captured by the parameter $\mu$, whose qualitative behaviour (i.e. whether it is real or imaginary) determines the physical properties of the solution.  Based on the broken Lorentz invariance of the backgrounds, the combinations that can appear a priori in $\mu$ are $\l \k$ and $\l T_+$. We will now give the explicit expression for $\mu$ in both the S-dual dipole and NHEMP truncations. This expression is obtained by reducing the fully coupled system of linear differential equations that describes the interrelated fluctuations of the metric, massive vectors and scalars above the black string backgrounds to a set of equations of the form \eqref{Master2} - one for each propagating degree of freedom in the theory. For details on this procedure, please consult appendix \ref{app:EOM}.

The S-dual dipole theory contains two propagating degrees of freedom, and thus $\mu$ takes on two different values. Remarkably, these have a very simple form, which is moreover temperature-independent,
\be
\mu_\pm = |1 \pm \frac{1}{2}\sqrt{1+\lambda^2 \kappa^2}| .\label{solsmu}
\ee
All modes have  $\mu^2 \geq 0$, which leads to power-law fall-offs. A plot for the lower branch $\mu_-^2$ as a function of $\k$ can be seen in figure 1. At the special values $\l \k = \pm \sqrt{3}$, we note that $\mu_-$ vanishes, and thus the perturbation is on the verge of becoming a travelling wave. We call this an evanescent travelling wave. 

\begin{center}
\DOUBLEFIGURE[hbt]{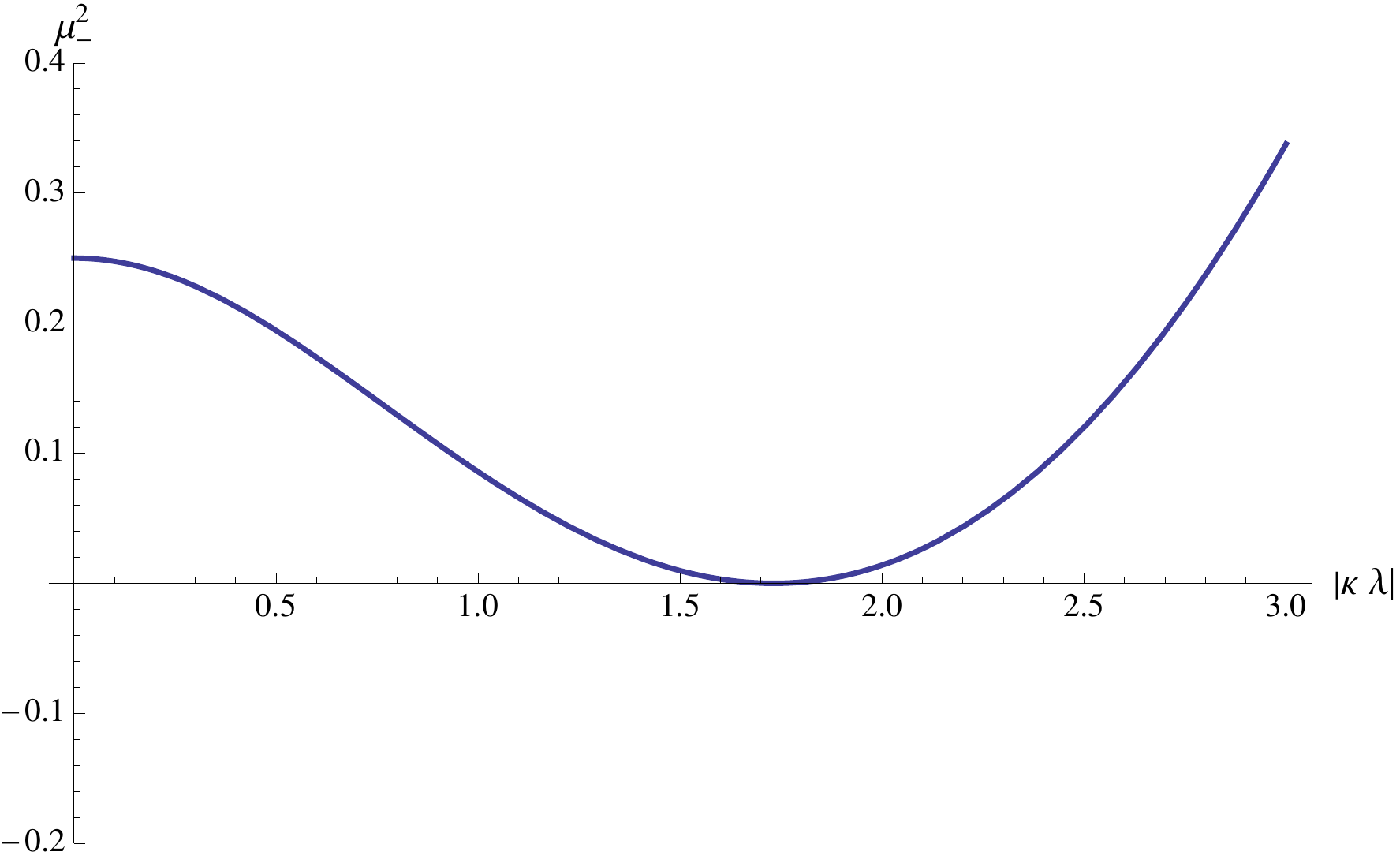,width=0.47\textwidth}{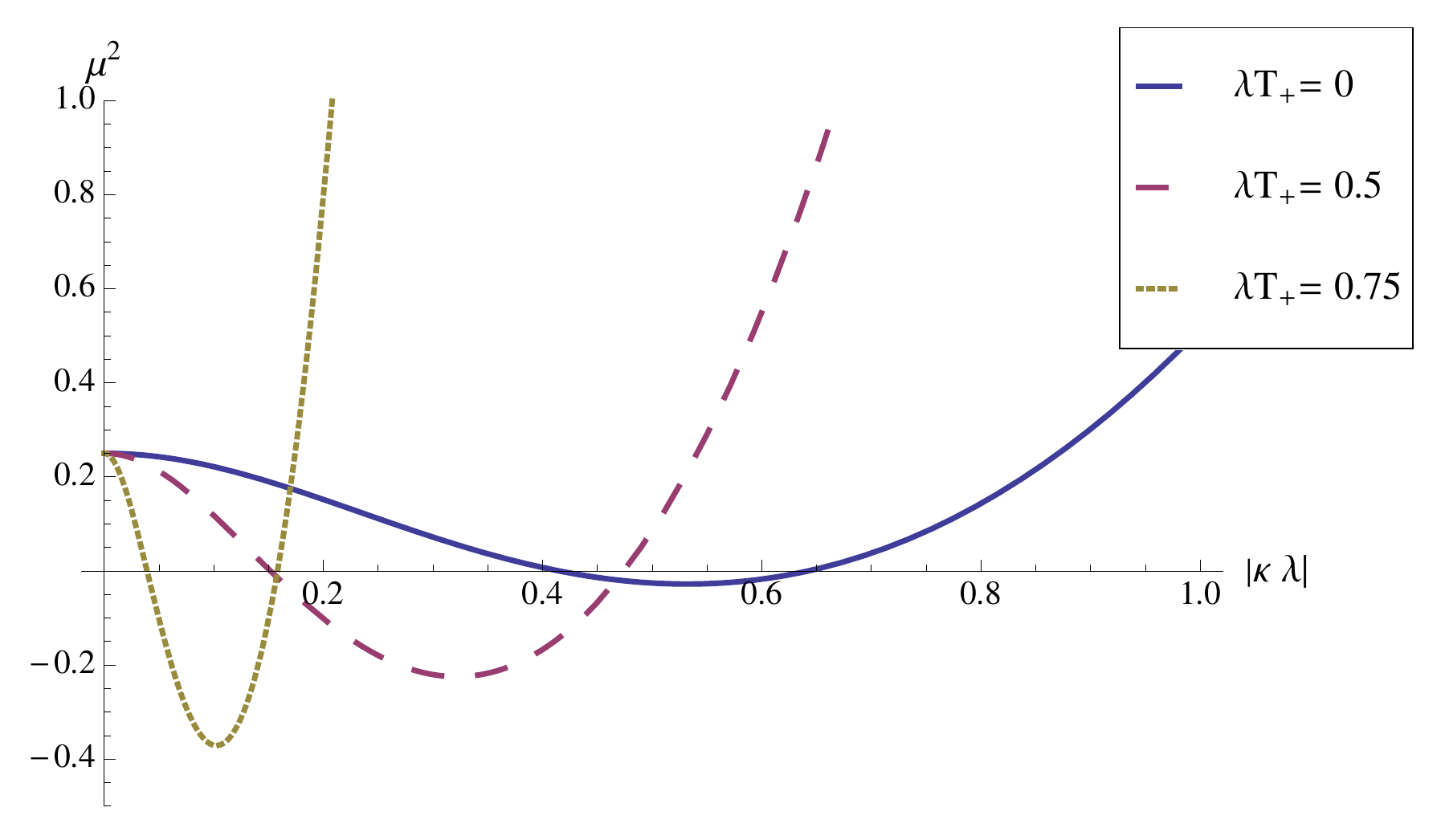,width=0.49\textwidth}{Spectrum (range of $\mu_-^2$) of the slower fall-off mode of the S-dual dipole theory as a function of the dimensionless  momentum $|\l \kappa|$. There is one evanescent travelling wave at $|\l \kappa| = \sqrt{3}$, where $\mu_- = 0$.\label{figdip}}{Spectrum (range of $\mu^2$) of the slower fall-off mode of the NHEMP truncation as a function of the momentum $|\l \kappa|$ at zero temperature  and two non-zero temperatures. There is always a finite range of momenta for which we obtain travelling waves with $\mu^2 <0$. \label{fignhek}}

\end{center}

\noindent For the NHEMP truncation, there are four degrees of freedom; accordingly, $\mu^2$ obeys a quartic equation, whose coefficients  depend on $\lambda \kappa$ and $\lambda T_+$. At zero momentum, the four solutions are $\mu =  \left\{\frac{1}{2},\frac{3}{2},\frac{3}{2},\frac{5}{2} \right\}$ and are independent of the temperature. At large momentum, all solutions asymptote to 
\be
\mu^2 \sim \frac{3+2 (\l T_+)^2 + 3 (\l T_+)^4}{(1-\l^2 T_+^2)^4} \l^2 \kappa^2 + \O(\l \kappa) \label{asymu}
\ee
Note that $\mu$ diverges as $\l T_+ \r 1$, i.e.  when $T_+$ is comparable to the non-locality scale of the dual field theory.
It would be interesting to understand the meaning of this divergence and why it occurs in the
  NHEMP truncation, 
  but not in the S-dual dipole theory.

The last three solutions of the quartic are monotonically increasing with momentum at any temperature
; nevertheless for the first one, for most temperatures, there is a finite range of momenta such that $\mu^2 < 0$. This situation has been depicted in figure \ref{fignhek} and  leads to travelling waves. Outside this range of momenta, the solution for $\mu$ is real, leading to power-law fall-offs. 
The explicit equation obeyed by  the function $\mu$  is given in \eqref{muf}.

The presence of travelling waves in the NHEMP truncation is rather worrisome, 
given that modes with $\mu \in i \mathbb{R}$ are usually associated with
 tachyons (i.e.  fields below the Breitenlohner-Friedman bound \cite{Breitenlohner:1982bm}), which are known to lead to instabilities.
The presence or absence of instabilities strongly depends on the asymptotic boundary conditions that one imposes, as was shown in \cite{Amsel:2009ev,Dias:2009ex,sss}. Let us briefly recall the results of these analyses. 

There are two natural boundary conditions to consider for travelling waves: (i) no flux passing through the spacetime boundary and (ii) purely outgoing flux passing through the boundary\footnote{When the warped AdS$_3$ spacetime is obtained from the near-horizon limit of an asymptotically flat black hole, a third natural boundary condition  is to require that there be no incoming flux from past null infinity in the asymptotically flat region, which in turn fixes the  ratio of incoming to outgoing flux  through the boundary of the near-horizon region to a definite value. This situation was considered e.g. in  \cite{Porfyriadis:2014fja,Hadar:2014dpa}.}.
The boundary condition (i) is the natural one to consider in order to have conserved symplectic flux, which implies that the phase space is self-contained and decoupled from external systems. When $\mu \in i \mathbb{R}$, the zero flux condition requires $|A|=|B|$ in \eqref{asyu}, which exactly balances the outgoing flux against the incoming one. Together with regularity in the interior of the spacetime (or ingoing boundary conditions at the coordinate horizon), this imposes a quantization condition on the allowed frequencies, $\om$, of the perturbation. It was shown in \cite{Amsel:2009ev} for NHEK (the proof can be easily repeated for the black string backgrounds) that when $T_+ > 0$, this quantization condition leads to a positive imaginary part for $\om$, which implies that the spacetime  has instabilities (exponentially growing modes in $u$). Nevertheless, when $T_+ =0, \, T_-=i$, \cite{Moroz:2009kv,sss} showed that no exponentially growing modes are present, and thus the vacuum is stable at linearized level\footnote{However, the spectrum of allowed $\omega$ is unbounded from below \cite{Moroz:2009kv}, which might lead to  cascading instabilities once interactions and/or quantum effects are included. }. A possible physical interpretation of this result has been given in \cite{Moroz:2009kv}. It would be very interesting to understand the fate of the instabilities present when $T_+ \neq 0$.

The second (ii) boundary condition is the one usually  imposed in order to study the late-time effects of an initial perturbation. It was shown in \cite{Dias:2009ex}
that under such boundary conditions, the near-horizon  geometry of the extreme Kerr black hole (NHEK) is stable, because travelling waves  have a discrete spectrum of complex frequencies, all of which have  negative imaginary part; thus, they represent quasi-normal modes.
The same analysis applies to our case, because the  equation we need to solve is the same. In the case of the BTZ black hole, it was shown in \cite{Birmingham:2001pj} that the quasi-normal modes can be mapped to the poles of the retarded Green's function of the dual CFT$_2$; a similar interpretation was suggested in \cite{Bredberg:2009pv} for NHEK. 
   Since flux is allowed to leak through the boundary, the near-horizon geometry is  not entirely decoupled from the exterior region, and thus it is natural to interpret the boundary field theory 
as being coupled to an external system. 



We thus conclude  that in the case of travelling waves, we have a choice between instabilities at finite temperature for boundary condition (i)   and having a non-conserved symplectic flux for (ii), neither of which sounds very appealing. Since the travelling waves only occur for a small range of momenta (see figure \ref{fignhek}), one could attempt to exclude them at the linear level. Nevertheless, 
 they are expected to reappear at non-linear level, due to interactions with the normal ($\mu \in \mathbb{R}$) modes. Since we do not yet know
what is the best physical interpretation of the travelling waves, we leave their treatment to future work and concentrate instead on the S-dual dipole theory.  

\subsection{Phase space for the S-dual dipole theory \label{sddipphsp}}

We would like to  construct a phase space that contains propagating modes in addition to the boundary gravitons of section \ref{sec:bnd}. We will only consider  Dirichlet boundary conditions for the non-bulk-propagating sector, since this is the case of interest for describing a theory with two sets of Virasoro descendants. For commodity, we shall denote by $\mathrm{T}_L = \{ \cL_{\xi_L} \phi^i\}$ and $\mathrm{T}_R= \{ \cL_{\xi_R} \phi^i\}$ the left/right-moving linearized boundary graviton modes\footnote{The reader should not confuse $\mathrm{T}_L,\mathrm{T}_R$ with the left/right-moving temperatures.} (here $\xi_{L/R}$ belongs to the left/right-moving Virasoro asymptotic symmetry algebra), by $\mathrm{T}= \{ \mathrm{T}_L,\mathrm{T}_R\}$ -  a generic linearized boundary graviton and by $\mathrm{ X} = \{ \delta \phi^i\}$ - a generic propagating mode. Here $\phi^i = g_{\mu\nu},\dots$ denote all the fields in the theory. 

As discussed in section \ref{symp}, there are several physical requirements that the symplectic form must satisfy in order to build a consistent phase space. One requirement is that it be normalizable
\be
\omega_{ru} = o(r^{-1}), \qquad \omega_{vr} = o(r^{-1}) \label{normom}
\ee
and conserved
\be
\omega_{uv}=o(r^0) .\label{consom}
\ee
The above relations must hold for all pairs of modes in the theory and around any background. For simplicity, we will restrict our analysis to linearized perturbations  around the black string backgrounds. Moreover, we will only consider  those with $T_- =0$ since, as we discussed in section \ref{setup}, a non-zero left-moving temperature can be induced by a coordinate transformation, and thus we believe that it will not be very hard to extend our conclusions to the general $T_- \neq 0$ case. 

The computation we perform is the following. We  consider pairs of  modes with momenta $\k_1$, $\k_2$ and  frequencies $\omega_1$, $\omega_2$, which can either be boundary gravitons ($\mathrm{T}$ modes, left or right), or propagating modes ($\mathrm{X}$ modes). We compute the total symplectic form for these modes \eqref{wisym},  including  a boundary contribution of the form $\boldsymbol d \boldsymbol \om_{\boldsymbol Y}$ \eqref{omY}, where $\boldsymbol Y $ is given by the expression \eqref{Ygen}. We will separately analyse the contributions from the $\mathrm{T} \mathrm{T}$, $\mathrm{X}\mathrm{X}$ and mixed $\mathrm{T}\mathrm{X}$ pairs of modes to the total symplectic form.

The symplectic form for the $\mathrm T \mathrm T$ modes was computed in section \ref{equiv}, where we have shown analytically that all divergences are removed by  the addition of the boundary counterterm $\boldsymbol Y$. We can indeed check explicitly that 
\bea
\omega_{ur}( \mathrm{T}_1,\mathrm{T}_2) = \O(r^{-2})\;, \quad \quad \omega_{vr}( \mathrm{T}_1, \mathrm{T}_2) = \O(r^{-2})\;,  \quad \quad \omega_{uv}( \mathrm T_1, \mathrm{T}_2) = \O(r^{-1})
\eea
and thus it satisfies the  requirements  \eqref{normom} - \eqref{consom}. We would now like to show that they also hold (or they can be made to hold) for the $\mathrm X \mathrm X$ and $\mathrm X \mathrm T$ pairs of modes.
%

 Let us now compute the symplectic form for the $\mathrm X \mathrm X$ pair of modes. We find
\be
\omega_{ur}( \X_1, \X_2) \sim   r^{-1-\mu_1 - \mu_2} \;,\;\;\;\;\;\;
\omega_{vr}( \X_1, \X_2)\sim  r^{-2-\mu_1-\mu_2},  \non
\ee
\be
\omega_{uv}(  \X_1, \X_2) \sim  r^{-\mu_1-\mu_2}.
\label{omx}
\ee
Since $\mu_{1,2} \geq 0$ 
for all momenta $\kappa_{1,2}$, the symplectic structure is both finite and conserved, except perhaps when
$\mu_1=\mu_2 = 0$, i.e. when  both modes are evanescent travelling waves. Nevertheless, one can check that for the particular values of $\k_i$ for which $\mu_i=0$, namely  $ \lambda \kappa_{1,2} = \pm \sqrt{3}$, the coefficients in front of the leading term of $\omega_{ur}$ and $\omega_{uv}$ vanish, and therefore no divergences are introduced.

Let us now compute the symplectic product between the propagating modes and the boundary gravitons. We parametrize the boundary gravitons in terms of the left/right-moving Virasoro generators and we work  in momentum space. Note that the left-moving Virasoro generators only carry left-moving momentum $\om$ (and have $\k=0$), whereas the right-moving generators have $\om=0$ and arbitrary right-moving momentum $\k$. Concretely, the generators are
%
\bea
\xi_L &=& e^{-i u \omega}\left[ \left(1-\frac{\omega^2 T_+^2}{8r^2} + \O(r^{-3})\right)\p_u + \left(\frac{\omega^2}{4r}+\O(r^{-2})\right)\p_v+ i \omega r \p_r \right], \\
\xi_R &=& e^{-i v \kappa}\left[ \left(\a_R \, \frac{\kappa^2}{4r}+\O(r^{-3}) \right) \p_u + \left(1+ \O(r^{-2})\right)\p_v+ i \kappa r \p_r \right],
\eea
where the term parametrized by $\a_R$ is a trivial diffeomorphism. Diffeomorphisms with $\a_R =1$ preserve 
the radial gauge for the untwisted BTZ metric $\hat g_{\mu\nu}$, whereas those with  $\a_R=0$ preserve the radial gauge $g_{rr}=g_{ru}=A_r=0$. We will keep $\a_R$ arbitrary, in order to ensure that our results are gauge-independent.

The mixed symplectic form between left-moving gravitons and propagating modes satisfies
\be
\omega_{ur}( \X, \mathrm{T}_L) = \O(r^{-\frac{5}{2}-\mu})\;, \qquad \qquad \omega_{vr}(\X, \mathrm{T}_L) = \O(r^{-\frac{7}{2}-\mu}),\non
\ee
\vspace{-7mm}

\be
\omega_{uv}( \X, \mathrm{T}_L) = \O(r^{-\frac{3}{2}-\mu})
\label{omtl}
\ee
for any $\a_R$. Since $\mu = \mu_{\mathrm{X}}\geq 0$, the symplectic structure is finite and conserved. For the mixed form involving right-moving gravitons, we find
\be
\omega_{ur}( \X, \mathrm{T}_R) = r^{-\frac{1}{2}-\mu} \, \omega_{ur}^\infty(\kappa_X,\kappa_T,T_+)e^{-i (\kappa_X+\kappa_T) v - i \omega u} +\O(r^{-\frac{3}{2}-\mu}) ,\nn\ee

\vspace{-3mm}

\be
\omega_{vr}(\X, \mathrm{T}_R) = \O(r^{-\frac{3}{2}-\mu}),\label{fall1}
\ee

\vspace{-3mm}

\be
\omega_{uv}( \X, \mathrm{T}_R) = r^{\frac{1}{2}-\mu} \, \omega_{uv}^\infty(\kappa_X,\kappa_T,T_+)e^{-i (\kappa_X+\kappa_T) v - i \omega u} +\O(r^{-\frac{1}{2}-\mu}),\nn
\ee
where $\mu = \mu_X$, $\om = \om_X$.  For all $\mu > 1/2$, the symplectic structure is finite and conserved. Nevertheless, for the range $0 \leq \mu \leq 1/2$, which occurs as shown in figure \ref{figdip}, both $\om_{ur} (\X,\mathrm{T}_R)$ and $\om_{uv} (\X, \mathrm{T}_R)$ violate the requirements \eqref{normom} - \eqref{consom}, leading to divergent symplectic  flux and norm. However, the expressions for $\om_{\mu\nu}^\infty$  satisfy
\be
(-\mu+\frac{1}{2})\, \omega_{uv}^\infty + i (\kappa_X+\kappa_T)\,\omega_{ur}^\infty = 0, \label{rel}
\ee 
which implies that the divergence\footnote{ In the case $\mu = 1/2$, \eqref{rel} implies that $\omega_{ur}^\infty  = 0$, and thus the symplectic structure is finite. The counterterm absorbs the finite flux proportional to $\omega_{uv}^\infty$.} can be absorbed into a supplementary boundary term of the form $\boldsymbol d \boldsymbol Y^{ct}$, where $\boldsymbol Y^{ct}$, when evaluated on the $\X \mathrm{T}_R$ sector, reads
\be
Y^{ct}_u = r^{-\mu + \frac{1}{2}} i (\kappa_X+\kappa_T)^{-1} \omega_{uv}^\infty e^{-i (\kappa_T+\kappa_X) v - i \omega u} +\O(r^{-\frac{1}{2}-\mu}) \label{yct}
\ee
\be
Y^{ct}_v = O(r^{-\frac{1}{2}-\mu})\;, \;\;\qquad Y^{ct}_{r} = O(r^{-\frac{3}{2}-\mu}).
\ee
The next step is to recognize $\boldsymbol Y^{ct}$ as a covariant expression. We propose
\bea
\boldsymbol Y^{ct} = A_\mu \delta f_1(U) dx^\mu + \eps_{\mu\nu\lambda}\delta (A^\nu) \p^\lambda f_2(U) dx^\mu \label{ans1}
\eea
for some functions $f_{1,2} (U)$, which also depend non-trivially on $\l \k$. Note that, by construction, the above counterterm can only contribute  to the symplectic product of $\X \mathrm{T}$ and $\X \X$ pairs of modes. By matching the divergent terms in \eqref{ans1} to those in \eqref{yct}, one can find an expression for $\p_U f_{1,2}$, evaluated on the background of interest, which does not depend on the gauge parameter\footnote{ This  does not prove that we obtained the unique counterm for any gauge, but at least in the gauges parameterized by $\a_R$, we find a unique consistent counterterm given by \eqref{ans1}.} $\a_R$. For example, around the null warped background, we find

\bea
\left. \frac{\p f_1}{\p U} \right|_{U=0}&=& i \frac{(1+2\mu_-)(-7+\kappa^2 \lambda^2 -4 \mu_-^2)}{\kappa \lambda (1-2\mu_-)},\non \\
\left. \frac{\p f_2}{\p U} \right|_{U=0}&=& i \ell \frac{5+18 \mu_- + (-3+2\mu_-)(\kappa^2 \lambda^2 -4\mu_-^2)}{2\kappa \lambda (1-2\mu_-)} ,\label{explf}
\eea
where $\mu_-$ has been given in \eqref{solsmu}. One may obtain the full solution for $f_{1,2} (U)$ by evaluating their derivative around all the other black string backgrounds and then integrating. The expression is divergent at $\kappa \rightarrow 0$ but there is no divergence at  $\kappa= 2\sqrt{2}$ which also corresponds to $\mu_-=\frac{1}{2}$.

The dependence of the denominators in \eqref{explf} on the momenta, and in particular the fact that they diverge as $\k \r 0$, indicates that the boundary counterterms are very non-local when expressed in position space. Similar non-local counterterms have been previously encountered   in  holographic renormalization for Schr\"odinger spacetimes \cite{Guica:2010sw}, 
and their appearance has been justified by the fact that  the dual theory  is non-local in the $v$ direction. These two types of counterterms are in fact related since, as shown in  \cite{Compere:2008us}, it is possible to 
 define the boundary term $\boldsymbol Y$ in the presymplectic structure in terms of the usual boundary counterterms that renormalize the on-shell action, in such a way that the final symplectic structure is finite. Here 
  we did not work out  the  counterterms, but we  simply derived this boundary term by requiring finiteness of the total symplectic structure. 


The boundary counterterm above has been engineered to remove the divergences in the mixed symplectic structure between propagating modes and right-moving boundary gravitons. Nevertheless, we need to check that no new divergences are introduced in the $\X \X$ and $\X \mathrm{T}_L$ cross products. 
%
It can be easily checked that the asymptotic behaviour of the  contribution  \eqref{ans1} to these products is still given by \eqref{omx}  and \eqref{omtl}. 
As before, the only potentially divergent terms have $\mu_1=\mu_2=0$, but one can again check that the overall coefficient vanishes. 
Therefore, no new divergences are introduced.


Besides finiteness, another important issue is the positivity of the symplectic form. For linearized perturbations around flat space/AdS with standard boundary conditions and uncharged matter, the original symplectic form (before any counterterms are added) is expected to be positive \cite{Ishibashi:2004wx}. For perturbations around backgrounds where non-trivial matter fields have been turned on, such as the warped AdS$_3$ background, positivity is not so clear.  In our case, positivity can in principle be  checked, since the perturbations are given by hypergeometric functions, which are known to satisfy the kind of completeness relations usually used in such computations. For the rest of our discussion, we will simply assume that the original symplectic form of the S-dual dipole theory around the warped AdS$_3$ backgrounds is positive definite for positive frequency perturbations. 

Even if the original symplectic form is positive definite, the boundary counterterms $\boldsymbol Y$ bring in a usually negative contribution (because they need to cancel a positive divergence), and can lead to non-positive norm directions in the final symplectic structure. This phenomenon was shown to arise for scalar fields in AdS with Neumann boundary conditions above the unitary bound \cite{Andrade:2011dg,Andrade:2011aa}. In the latter analysis, null states were found at zero boundary momentum where infrared divergences appeared. In our analysis, infrared divergences at zero boundary momentum $\kappa = 0$ also appear but the structure of counterterms crucially differ from \cite{Andrade:2011dg,Andrade:2011aa}. In the following, we will show that if the original symplectic form of the S-dual theory is positive definite, as we assumed above, then this fact is not affected by the addition of the boundary counterterms.

First, we showed that the final symplectic form in the $\mathrm{T} \mathrm{T}$ sector is identical to that of three-dimensional Einstein gravity in AdS$_3$, and thus is positive definite. In the $\X \mathrm{T}$ sector, the initial symplectic form is divergent, and thus one may fear that upon the subtraction of $\boldsymbol Y^{ct}$, the final answer may not be positive definite. Nevertheless, note that by \eqref{chom1} both the original symplectic form and the counterterm contribution are given by  total divergences, which cancel each other in the final answer, yielding zero.  It implies that the X modes and T modes are orthogonal\footnote{Note that, while the propagating modes are not charged under the Virasoro symmetries at the linear level, they are expected to carry non-trivial charges at second order in perturbation theory, after backreaction effects are taken into account. }. Finally, in the $\X \X$ sector, both the original and the renormalized symplectic form are finite for $\mu >0$. Nevertheless, the boundary contributions can be reduced to a boundary integral over $Y_{u} \sim r^{-\mu_1-\mu_2}$, which vanishes as $r \r \infty$ if $\mu_i >0$.  The contributions with $\mu =0$ need to be treated separately, and we checked explicitly that they all vanish.

In summary, we showed that one can define a finite and conserved symplectic structure for the S-dual dipole theory, at the linear level  around the warped AdS$_3$ spacetimes, which includes two sets of Virasoro generators. We also showed that this symplectic form will be positive definite if the symplectic form of the theory before the inclusion of the right-moving Virasoro generators was positive definite.  If the latter is true, then a unitary quantum conformal field theory might exist, which is dual to the warped AdS$_3$ spacetime.

\subsubsection*{Further comments}

Note that our analysis above depends rather little on the specific details of the theory we study, since most of the falloffs in the symplectic form are simply dictated by the asymptotic falloffs  \eqref{asyu} - \eqref{fall} of the solution. Thus, our conclusions above apply to all theories whose spectrum only contains modes with $\mu>0$.

Even when travelling waves with $\mu \in i \mathbb{R}$ are allowed, we expect to be able to make the symplectic form finite. The first step is to impose the zero-flux boundary condition for the $\X \X$ pairs of modes at the warped AdS$_3$ boundary which, as previously discussed, imposes a certain relation between the asymptotic coefficients $A$ and $B$ in \eqref{asyu}. Disregarding the instabilities triggered by this boundary condition (which should be matched by a corresponding instability of the dual field theory), we proceed by requiring that the symplectic product of the $\X \mathrm{T}$ pairs of modes be finite. Since Re$\,\mu=0$, we find divergences for the right-moving gravitons for all the travelling waves. These divergences can again be absorbed by a boundary counterterm $\boldsymbol d \boldsymbol Y^{ct}$, which leaves us with a finite contribution. 
 Note though that now the boundary  counterterms can contribute non-trivially to the $\X \X$ symplectic product, and thus positivity needs to be re-checked explicitly.

Thus,  the symplectic form can always be made finite at linearized order around the warped black string backgrounds; 
this follows from the arguments of \cite{Compere:2008us} and the assumption that there is a good variational principle. 
 An interesting question is whether this result extends to non-linear order. For this we need to understand the fall-offs of the modes order by order in perturbation theory. Such an analysis was carried out in e.g. \cite{vanRees:2011fr,Dias:2011ss}, who showed that the same leading asymptotic behavior is present at higher orders in perturbation theory and certain subleading terms are introduced. If our theories follow the same logic, the form of the counterterms required to make the symplectic structure finite will therefore be of the same form as the one discussed, with possibly different coefficients $f_1(U)$, $f_2(U)$, which can be computed in perturbation theory. Again, travelling waves introduce more features and must be treated separately. We conclude that theories whose spectra only contain perturbations with $\mu \geq 0$, such as the S-dual dipole theory, should have a well-defined phase space even at non-linear level, which bears the action of two Virasoro symmetries. Whether this phase space can lead to a quantum theory that does not contain ghosts depends on the positivity properties of the symplectic form.



\section{Discussion}

This paper presents two main results. Our first result is that the symplectic form of certain theories admitting warped AdS$_3$ solutions, when restricted to a non-dynamical subsector, is exactly equal to the symplectic form of pure Einstein gravity in three dimensions. 
Using this equivalence of symplectic structures, we are able to map all known consistent  boundary conditions for AdS$_3$ to consistent boundary conditions for warped AdS$_3$, and to construct phase spaces for warped AdS$_3$
whose symmetries consist of either  two copies of the Virasoro algebra or a Virasoro $\times$ Ka\v{c}-Moody algebra. We note that the construction of two Virasoro asymptotic symmetry algebras in warped AdS$_3$ is new, even though their existence was to be expected in view of the results of the holographic analyses of \cite{fg,decrypt}. 

Our second main result is the proof that the Virasoro $\times$ Virasoro symmetry can be extended to the entire, dynamical, linearized phase space in theories whose spectrum does not contain travelling waves, such as the S-dual dipole theory. This implies that, at least at semi-classical level, the field theory dual to such theories in warped AdS$_3$ is a two-dimensional conformal field theory. This result is rather surprising, given our current understanding of the holographic dual: it can be either defined via an irrelevant but yet exactly marginal (with respect to the left-moving scaling symmetry) operator deformation, which breaks all the right-moving conformal symmetries \cite{Guica:2010sw}, or it can be (roughly) understood in terms of the dipole deformation \cite{Bergman:2000cw} of the D1-D5 gauge theory, flown to the infrared, which again does not seem to leave any space for a right-moving Virasoro symmetry.  It would be thus interesting to understand whether these symmetries extend beyond the supergravity spectrum to the string states\footnote{For previous attempts at finding these symmetries in the string spectrum see \cite{Azeyanagi:2012zd}. }, as well as to non-perturbative level.  

An important question is whether the final symplectic structure on the phase space is positive definite. 
If true, this would imply that the CFT dual to warped AdS$_3$ is unitary. Even though we did not compute the symplectic structure in the bulk, we argued that the ghosts appearing in previous treatments of modified symplectic structures with boundary counterterms \cite{Andrade:2011dg,Andrade:2011aa} do not arise here. It would be very interesting to check positivity of the symplectic form for these theories.

Our construction of the non-dynamical sector of the  warped AdS$_3$
phase space has the interesting feature that it is \emph{local}, for both Dirichlet and mixed boundary conditions in the auxiliary AdS$_3$ space-time. 
 This constuction crucially uses the decomposition of the holographic stress-energy tensor in terms of the stress-energy tensor of a chiral boson with exponential potential. While this decomposition naturally arises in the Hamiltonian reduction of pure Einstein gravity in terms of Liouville theory in Chern-Simons formalism \cite{Coussaert:1995zp}, it has not been used to the same extent in the metric formalism.  It would be interesting to investigate whether this observation can be used to reformulate in a local fashion the non-localities arising in the presence of sources as well, given that sources in Einstein gravity are encoded in a conformally flat boundary metric, whose Weyl factor is known to lead to a boundary Liouville theory \cite{Polyakov:1987zb}.


An interesting difficulty that we encountered in our analysis was the treatment of the travelling waves. Such modes are generically present in the near-horizon of extreme black holes, but their holographic description is very poorly understood (see however \cite{Moroz:2009kv}). When travelling waves are present,  either the symplectic structure is not conserved, or instabilities are present at finite temperature
. It would be very interesting to understand what sort of field theory can reproduce this behavioural pattern on the gravity side.

At the technical level, our analysis  relies heavily on the fixation of ambiguities in the symplectic structure by adding suitable boundary terms. While we find boundary counterterms which cancel all divergences, it would be instructive to have  a more fundamental derivation of these  terms - for example in the framework of holographic renormalization - 
as boundary terms appearing in the variation of the counterterms\footnote{In asymptotically AdS spacetimes, such boundary terms only matter for non-Dirichlet boundary conditions for which the boundary fields fluctuate  \cite{Compere:2008us}. By contrast, in warped AdS spacetimes  ``Dirichlet'' boundary conditions allow the leading order field to fluctuate, and therefore these boundary counterterms  become important even in this case.} \cite{Compere:2008us} (see also \cite{Skenderis:2008dg}). More generally, it would be interesting to fully compare holographic renormalization for warped AdS$_3$ spacetimes  with the covariant phase space techniques that we used herein, and show that the
conserved charges computed using the two formalisms agree, as was established  in \cite{Hollands:2005wt,Hollands:2005ya,Papadimitriou:2005ii} for AdS. 

%
%

Finally, a very interesting future direction would be to understand how to extend our analysis to the near-horizon region of the extreme four-dimensional Kerr black hole, as well as to all the extremal black holes to which the original Kerr/CFT analysis applies. There are two main problems to surmount: one is the lack of dynamics of these near-horizon geometries; nevertheless, an already interesting first step would be  to build  the non-dynamical version of the phase space, as in section \ref{sec:bnd} above. The other problem is that in our analysis, the massive vector $A_\mu$ played an essential role in the construction of the counterterms that remove the divergences in the symplectic structure, but no such  vector field  seems to be readily available e.g. for the four-dimensional extremal Kerr black hole.
We thus leave to future work the construction of both Virasoro asymptotic symmetries for more realistic black holes.



\acknowledgments

We are grateful to  Tom Hartman, Simon Ross, Kostas Skenderis, Andrew Strominger, Marika Taylor and especially Don Marolf for useful conversations.
G.~C. is a Research Associate of the Fonds de la Recherche Scientifique F.R.S.-FNRS (Belgium) and is currently supported by the ERC Starting Grant 335146 ``HoloBHC''. He gratefully thanks the University of Amsterdam for its hospitality during the Amsterdam string workshop 2014.
 The work of M.~G. was supported in part by the DOE grant DE-SC0007901 and by the National Science Foundation Grant No. PHYS-1066293 and the hospitality of the Aspen Center for Physics.  M.~J.~R. was supported by the European Commission - Marie Curie grant PIOF-GA 2010-275082.
 
\appendix

\section{Details of the computations}

\subsection{The linearized solution for  boundary gravitons in radial gauge \label{linsol}}

In this section, we would like to present the  explicit linearized solution for the non-bulk-propagating modes in the case when $\hat g_{\mu\nu}$ obeys Dirichet boundary conditions and we work in radial gauge (unlike in section \ref{excon}), in which the metric components have a simple holographic interpretation. For simplicity, we will concentrate on the example of the null warped background, for which $U=0$, and thus  $\hat A^2=0$. We will also be working in terms of the $u,v,r$ coordinates used throughout the paper, with the exception of sections \ref{excon} and \ref{dnbc}.
We take the background $\hat g_{\mu\nu}$ metric to  be Poincar\'e AdS$_3$
\be
d\hat s^2 = \ell^2 \left( 2 r du dv + \frac{dr^2}{4r^2} \right). \label{poincads}
\ee
A solution for $A_\mu$ satisfying the requirements \eqref{consAb} with $U=0$ is
\be
A = \l \ell r u
\ee
for some parameter $\l$. All other solutions to \eqref{consAb} with $U=0$ are related to it by coordinate transformations.

We would now like to consider linearized perturbations around the background \eqref{poincads}, $\hat h_{\mu\nu}$, which satisfy the equations of motion \eqref{eqhatg}, linearized around the background. In radial gauge $(\hat h_{\mu r} =0)$, the most general linearized solution is simply given by\footnote{In section \ref{sddipphsp}, we will also need the solution for $\hat h_{\mu\nu}$ when the background $\hat g_{\mu\nu}$ metric is BTZ with $T_+ \neq 0, T_-=0$. It is still parametrized by two holomorphic functions as
$$ \hat h_{uu} = \ell^2 F(u) \;, \;\;\;\;\; \hat h_{vv} = \ell^2 G(v) \;, \;\;\;\;\; \hat h_{uv} = \frac{ T_+^2 F(u)  \ell^2
}{4r}.
$$
 }
\be
\hat h_{uu} = \ell^2 F(u) \;, \;\;\;\;\; \hat h_{vv} = \ell^2 G(v) \;, \;\;\;\;\; \hat h_{uv} =0
\ee
where $F(u)$ and $G(v)$ are arbitrary functions of their respective arguments and are understood to be proportional to a small parameter. The solution for $A_\mu$ satisfying \eqref{consAb}, expanded to linear order, is
\be
A= \l \ell\, \left(1+ 2\, \e_G(v) \right) r du + \half \, \l \ell \, G(v) dv - \frac{\l \ell}{2r} \,\e'_G(v) dr
\ee
We have introduced a new function $\e_G(v)$, which satisfies
\be
\e_G''(v) \equiv G(v) 
\ee
%
%
This notation is somewhat similar to the one in \eqref{EQ1}, up to a factor of $\l$. The above equation shows very explicitly  that, given $\hat g_{\mu\nu}$, the solution for $A_\mu$ is non-local in terms of the boundary data, because in order to obtain $\e_G(v)$, one needs to integrate $G(v)$ twice. Constructing $g_{\mu\nu}$ using \eqref{defgh} and discarding all terms quadratic in the perturbation,  we find 
\be
\frac{ds^2}{\ell^2} =\left[ F(u) -  \l^2 r^2  (1+ 4 \e_G) \right]  du^2 +  r (2-\l^2 G) du dv+ \l^2 \e'_G du dr + G(v) dv^2 + \frac{dr^2}{4r^2}
\ee
Since, when Dirichlet boundary conditions are imposed, the functions $F(u), G(v)$ are allowed to fluctuate, we see from the above expression that the most divergent term in $g_{\mu\nu}$ (proportional to $r^2 du^2$) is allowed to fluctuate, whereas the subleading terms, such as the boundary metric in $\hat g_{\mu\nu}$, are held fixed. This is certainly a peculiarity of the solutions to the equations of motion in this space-time, but note that it matches perfectly with the boundary conditions \cite{Guica:2008mu} that were originally imposed on metric fluctuations in NHEK. 

The diffeomorphisms that leave the asymptotic AdS$_3$ metric invariant are
\be
\xi_{AdS}(\a) = \left( f(u) - \frac{\a_R}{4r}\, g''(v)\right) \p_u +  \left( g(v) - \frac{\a_L}{4r}\, f''(u)\right) \p_v - [f'(u) + g'(v)]\, r \, \p_r.
\ee
The terms proportional to $\a_{L,R}$ correspond to trivial diffeomorphisms in AdS$_3$, i.e. they do not contribute to the asymptotic charges. The choice $\a_{L,R}=1$ is nevertheless singled out, because if one acts with this diffeomorphism on the pure Poincar\'e AdS$_3$ metric one obtains a perturbation that is in radial gauge, with
\be
F(u) = -\half\, f'''(u) \;, \;\;\;\;\; G(v) = -\half\, g'''(v).
\ee
Using the above explicit expressions, one can study the divergences in the symplectic form $\boldsymbol \om$ and the conserved charges, and see explicitly how these divergences are completely removed by the introduction of the one-form $\boldsymbol Y$ in \eqref{defTB}. We leave this exercise to the interested reader.

%

\subsection{Linear perturbations in the NHEMP truncation}
\label{app:EOM}

The equations describing linearized perturbations around the null warped and black string backgrounds form a rather complicated set of coupled linear differential equations, which needs to be solved in a particular order  to be able to find the solution. The case of linearized perturbations around the black string backgrounds with general $T_\pm$ in the NHEMP truncation is particularly involved; the aim of this appendix is to present an
 algorithm for solving this system. Using it (and with the help of Mathematica), one can derive the hypergeometric equation \eqref{Master2}, the fall-off behavior \eqref{fall} and the expression for the parameter $\mu$ as a function of the right-moving temperature and the momentum.

The first step is to partially gauge fix the metric perturbation $h_{\mu\nu}$, by imposing $h_{r\mu}=0$. All fields are expanded in Fourier components, as explained in the main text. 
 We denote the eleven $r$-dependent linearized fields as $\phi^i = \{ h_{ab} , \d A^1_\mu ,\d A^2_\mu,\d U_1,\d U_2 \}$, $a=\{u,v\}$, $\mu=\{r,u,v\}$. The equations of motion are coupled ODEs for these eleven fields. Let us present a convenient way to solve them. 

We denote by $E_{\mu\nu}$ Einstein's equations \eqref{eq1}, by  $E^{1,2}_\mu$ the massive vector equations \eqref{eq2} for $A_1$ and $A_2$  and by $E^{1,2}$ the scalar field equations \eqref{eq3}-\eqref{eq4} for $U_1$ and $U_2$. These are 14 equations in total, but several are redundant.   Let us denote all equations as $E_\alpha = \{ E_{\mu\nu},E^{1,2}_\mu,E^{1,2}\}$. In the following, one can read $E_\alpha \Rightarrow \phi^i$ as solving $E_\alpha = 0$ for the field $\phi^i$. The algorithm is given by

\vspace{0.5 cm}
\begin{minipage}{0.5 \textwidth}
\begin{enumerate}
\item $E^2_r \Rightarrow \d A_u^2$
\item $E^1_r \Rightarrow \d  A_r^2$
\item $r E^2_v - T_+^2 E^2_u \Rightarrow  \d A_r^1$
\item $r E_{vr}-T_+^2 E_{ru} \Rightarrow \p_r h_{uv}$
\end{enumerate}
\end{minipage}
\begin{minipage}{0.5\textwidth}
\begin{enumerate}
\item[5.] $E_{vv}+\# T_+^2 E_{ur}+\# T_+^2 E_{vr}\Rightarrow \p_r^2 h_{vv}$
\item[6.] $E_{uv}+\# T_+^2 E_{ur}+\#T_+^2 T_-^2 E_{ur} \Rightarrow h_{uu}$
\item[7.] $E_{ur} \Rightarrow  \d A_u^1$
\item[8.] $E^1$ and $E^2$ $\Rightarrow  \d A_v^1$ and $ \d A_v^2$ 
\end{enumerate}
\end{minipage}
\vspace{0.1 cm}


\noindent where $\#$ denotes an appropriate quotient of polynomials of $r$ with $T_\pm,\kappa,\lambda$ dependent coefficients, which is chosen so as to eliminate the derivatives of the field that one is solving for (the equation to be solved is then algebraic and not differential). At the end of step 8, we find that $r E^1_v-T_+^2 E^1_u=0$. The residual linearized diffemorphisms are three linearly independent solutions which precisely correspond to the three integration constants that one would obtain by integrating the residual differential equations (in steps 4 and 5 above) for $\p_r^2 h_{vv}$ and $\p_r h_{uv}$. The last independent equations are $E^1_u$ and $E^2_u$. One can check that they do not depend upon $h_{vv}$, $\p_r h_{vv}$ and $h_{uu}$ but only upon $U_{1,2}$ and their derivatives. One can then reduce them to an 8th order ODE for $U_1$ as
\vspace{0.5 cm}

\begin{minipage}{0.5 \textwidth}
\begin{enumerate}
\item[9.] $E^2_u \Rightarrow \p^4 U_2$
\item[10.] $E^1_u \Rightarrow \p^2 U_1$
\end{enumerate}
\end{minipage}
\begin{minipage}{0.5 \textwidth}
\begin{enumerate}
\item[11.] $E^2_u \Rightarrow  U_2$
\item[12.] $E^1_u \Rightarrow \p^8  U_1$
\end{enumerate}
\end{minipage}
\vspace{0.1 cm}

\noindent The final equation is a 8th order ODE for $U_1$, which factorizes into four sets of second order ordinary differential equations of precisely the form \eqref{Master2}, which only differ by the value of the parameter $\mu^2$. This parameter obeys a quartic equation, whose coefficients depend upon $\lambda T_+$ and $\lambda \kappa$. Denoting by $\lambda\kappa = \hat\kappa$, $\lambda T_+ = t$, the quartic equation is 
\bea
0&=&\mu^8-\frac{(4 \hat \kappa^2 (3 t^4+2 t^2+3)+11 (t^2-1)^4)}{(t^2-1)^4}  \mu^6 + \nn \\
&&   +\frac{ (48 \hat \kappa^4 (3 t^4+2   t^2+3)^2+8 \hat \kappa^2 (79 t^4+42
   t^2+79) (t^2-1)^4+287   (t^2-1)^8)}{8(t^2-1)^8}  \mu^4 - \nn\\
  &&   -\left( \frac{ (64 \hat \kappa^6 (3     t^4+2 t^2+3)^3+16 \hat \kappa^4 (3 t^4+2   t^2+3) (59  t^4+18 t^2+59)  (t^2-1)^4}{16   (t^2-1)^{12}} +\right. \nn\\
&& \left.  + \frac{4 \hat \kappa^2 (373 t^4-114   t^2+373) (t^2-1)^8+639   (t^2-1)^{12})}{16   (t^2-1)^{12}}   \right)\mu^2\nn \\
   && +\frac{256 \hat \kappa^8 (3  t^4+2t^2+3)^4+256 \hat \kappa^6 (3   t^4+2 t^2+3)^2 (13 t^4-2
   t^2+13) (t^2-1)^4}{256   (t^2-1)^{16}}+\nn\\
&&   +\frac{32 \hat \kappa^4 (167  t^8-2636 t^6-3318 t^4-2636   t^2+167) (t^2-1)^8}{256   (t^2-1)^{16}}+\nn\\
   && +\frac{-720 \hat \kappa^2
   (19 t^4+66 t^2+19)   (t^2-1)^{12}+2025 (t^2-1)^{16}}{256   (t^2-1)^{16}}. \label{muf}
\eea
The equation is singular at $\lambda T_+ = 1$, which indicates that something special is happening in the dynamics of the NHEMP truncation when the temperature is equal to the inverse dipole length $1/\lambda$. This property remains to be understood. For large $ \hat \kappa$, the four branches asymptote to \eqref{asymu}.
One can also check numerically that all solutions for $\mu^2$ are real in the range $0 \leq t < 1$, $0 \leq \hat \kappa \leq 4$. Given an asymptotic solution to \eqref{Master2}, one can then reconstruct the asymptotic behavior of all fields by following the results of the algorithm backwards (reconstructing $U_2$, then $A_v^2$, etc). One finds \eqref{fall}.


\providecommand{\href}[2]{#2}\begingroup\raggedright\endgroup

\end{document}